\documentclass[natbib,sn-basic]{sn-jnl}

\def\Rm{\mbox{\rm Re}_{\rm M}}
\def\Rmc{\mbox{\rm Re}_{\rm M}^{\rm crit}}
\def\Rey{\mbox{\rm Re}}
\def\Pm{\mbox{\rm Pm}}

\newcommand{\avg}[1]{\ensuremath{\langle #1 \rangle}}

\jyear{2023}%

\usepackage{natbib}

\begin{document}

\title[Small-scale dynamos]{Small-scale dynamos: From idealized models to solar and stellar applications}

\author[1]{\fnm{Matthias} \sur{Rempel}}\email{rempel@ucar.edu}
\equalcont{These authors contributed equally to this work.}

\author[2]{\fnm{Tanayveer} \sur{Bhatia}}\email{bhatia@mps.mpg.de}
\equalcont{These authors contributed equally to this work.}

\author[3]{\fnm{Luis} \sur{Bellot Rubio}}\email{lbellot@iaa.es}
\equalcont{These authors contributed equally to this work.}

\author*[2,4,5]{\fnm{Maarit J.} \sur{Korpi-Lagg}}\email{maarit.korpi-lagg@aalto.fi}
\equalcont{These authors contributed equally to this work.}

\affil*[1]{\orgdiv{High Altitude Observatory}, \orgname{National Center for Atmospheric Research}, \orgaddress{\street{P.O. Box 3000}, \city{Boulder}, \postcode{80307}, \state{Colorado}, \country{USA}}}

\affil[2]{\orgdiv{Department of Sun and Heliosphere}, \orgname{Max Planck Institute for Solar System Research}, \orgaddress{\street{Justus-von-Liebig-Weg 3}, \city{G\"ottingen}, \postcode{37077}, \country{Germany}}}

\affil[3]{\orgdiv{Instituto de Astrof\'{\i}sica de Andaluc\'{\i}a}, \orgname{CSIC}, \orgaddress{\street{Glorieta de la Astronom\'{\i}a s/n}, \city{Granada}, \postcode{18008}, \country{Spain}}}

\affil[4]{\orgdiv{Department of computer science}, \orgname{Aalto University}, \orgaddress{\street{Konemiehentie 2}, \city{Espoo}, \postcode{00076}, \country{Finland}}}

\affil[5]{\orgdiv{Nordic Institute for Theoretical Physics}, \orgname{KTH Royal Institute of Technology and Stockholm University}, \orgaddress{\street{Hannes Alv\'ens v\"ag 12}, \city{Stockholm}, \postcode{10691}, \country{Sweden}}}

\newcommand{\mr}[1]{\textcolor{magenta}{MR: #1}}
\newcommand{\mjk}[1]{\textcolor{blue}{MJK: #1}}
\newcommand{\lb}[1]{\textcolor{red}{LB: #1}}

\abstract{
In this article we review small-scale dynamo processes that are responsible for magnetic field generation on scales comparable to and smaller than the energy carrying scales of turbulence. We provide a review of critical observation of quiet Sun magnetism, which have provided strong support for the operation of a small-scale dynamo in the solar photosphere and convection zone. After a review of basic concepts we focus on numerical studies of kinematic growth and non-linear saturation in idealized setups, with special emphasis on the role of the magnetic Prandtl number for dynamo onset and saturation. Moving towards astrophysical applications we review convective dynamo setups that focus on the deep convection zone and the photospheres of solar-like stars. We review the critical ingredients for stellar convection setups and discuss their application to the Sun and solar-like stars including comparison against available observations.}


\maketitle

\section{Introduction}\label{sec:intro}
The Sun is the only star that can be scrutinized in detail at high spatial and temporal resolution.
Observations show that magnetic fields are ubiquitous in the quiet Sun---the areas of the solar surface away from active regions and the enhanced network. They cover the whole solar surface, all the time, irrespective of the phase of the solar cycle. Quiet Sun (QS) fields can be classified into network and 
internetwork (IN) fields. The former are relatively strong and vertical, and occupy the outer boundaries of supergranular cells. The latter are weaker and highly inclined, and can be found in the cell interiors. In high-resolution observations, both fields are detected as magnetic flux concentrations of opposite signs organized on subarcsecond scales. Observations of quiet Sun magnetism have provided over the past few decades growing support for a dynamo process that operates in the IN independently from the large-scale dynamo (LSD)
responsible for the solar cycle. We provide a detailed review of solar observations in Sect.\ \ref{sect:observations}. 

The theoretical concept of a small-scale dynamo instability (hereafter SSD), independent of the presence of symmetry-breaking effects, such as helicity, dates back to \cite{K68}. As identified early on, the magnetic Prandtl number, $\Pm=\nu/\eta$, where $\nu$ is the kinematic viscosity and $\eta$ the magnetic resistivity, plays a critical role for 
SSD action and we provide a more detailed account of the theoretical concepts in Sect.\ \ref{sect:theory}. It was first suggested by \cite{1993A&A...274..543P} that this dynamo instability may be the origin of the weak IN fields. 

Numerical simulations of SSD started with convective setups and solar surface simulations with 3D radiative transfer at $\Pm$ around one, and have recently been extended towards the numerically more challenging lower $\Pm$-regime. The evolution of numerical models is described in Sect.\ \ref{sect:models}.
We focus on investigations of the role of $\Pm$ during kinematic and saturated phases (Sect.~\ref{subsect:Pm}), deep
convection setups (Sect.\ \ref{subsect:deepconvection}), surface convection setups (Sect.\ \ref{sect:surface}) and the possibility of a SSD in the radiative interior of solar-like stars (Sect.\ \ref{subsect:radiative}). Section \ref{sect:stellar} provides an overview of recent applications to solar-like stars. Here we focus on the effects of small-scale magnetism on stellar structure, the basal chromospheric flux and contributions to short-term stellar variability that has to be considered for exoplanet detection. We conclude the review with an outlook in Sect.\ \ref{sect:outlook}.

\section{Solar observations}\label{sect:observations}
Because of their abundance, quiet Sun fields are important
contributors to the flux budget of the solar photosphere. The total
longitudinal magnetic flux of the quiet Sun has been estimated to be
$8 \times 10^{23}$~Mx at any time \citep{gosic_thesis}, similar to the
total flux carried by active regions at solar maximum \citep[$\sim 6
\times 10^{23}$~Mx;][]{2011ApJ...731...37J}. Network fields
contribute about 80-85\% of this flux, while the IN supplies the
remaining 20-15\% \citep{1995SoPh..160..277W, gosic_thesis}. However,
the IN is extremely dynamic and evolves very rapidly, with flux
appearance rates from 120 to 1100~Mx~cm$^{-2}$~day$^{-1}$
\citep[e.g.,][]{2016ApJ...820...35G, 2017ApJS..229...17S} 
that surpass those of active regions by orders of magnitude 
\citep[0.1~Mx~cm$^{-2}$~day$^{-1}$;][]{1994SoPh..150....1S}. 
IN fields are also the main contributors to the total energy budget 
of the solar photosphere \citep{2004Natur.430..326T, Rempel:2014:SSD}. 

Pushed and dragged by granular and supergranular convective flows,
quiet Sun magnetic fields undergo frequent interactions between them
and with other pre-existing fields, particularly in the IN. Such
interactions are believed to trigger magnetic reconnection events at
different heights in the atmosphere, releasing energy and contributing
to the heating of the chromosphere and corona, both locally and
globally. This role has been recognized only recently, as high
resolution observations became available both from ground and from
space. Quiet Sun fields also contribute to atmospheric heating by
channeling waves from the photosphere to higher atmospheric layers 
\citep{2023LRSP...20....1J}.

To understand these fields we must determine their properties and
origin. From an observational point of view, they turn out to be very 
different from active region fields. For example, they are weaker, 
more dynamic, and do not seem to follow the solar cycle. Therefore, 
it is likely that their origin is also different from the 
LSD-related active regions and magnetic fluctuations originating from this 
component through tangling by turbulence.
Cascading from large to small scales by active region decay seems not 
to be viable because of the low flux emergence rate of active regions 
and the lack of solar cycle variations. An SSD was proposed by 
\cite{1993A&A...274..543P} as the origin of the weak IN fields. 
SSD simulations are indeed able to reproduce the main 
statistical features of the quiet Sun magnetism, but important 
questions remain open, as for example whether or not they can explain 
the spatial distribution of the flux, the field strength and 
field inclination distributions, or the flux emergence processes
observed in the quiet Sun.

Conversely, by characterizing the properties, dynamics and
temporal evolution of quiet Sun fields observationally, it should be
possible to constrain SSD models and help set the most
appropriate physical ingredients such as 
boundary conditions and $\Pm$ regimes.
In particular, a detailed characterization of
the flux emergence process in granules and intergranular lanes using
high-resolution observations is needed to validate the mechanisms
suggested by numerical simulations.

\subsection{Diagnostic techniques}
Our current understanding of quiet Sun fields has been gained through the
interpretation of polarimetric measurements of spectral lines formed
in the solar photosphere. The most precise inferences are obtained
by inverting the radiative transfer equation \citep{2016LRSP...13....4D}, 
although in some cases also the
polarization signals themselves are employed to study the magnetism of
the quiet Sun, particularly when they are very weak.

Both the Zeeman and Hanle effects are used, as the polarization
signatures they generate depend on the properties of the vector magnetic
field. Zeeman measurements provide spatially resolved observations, but
are affected by possible cancellation of the opposite-polarity fields
that may coexist in the resolution element. Hanle measurements do not
suffer from these cancellation effects, but in turn they are spatially 
unresolved.

The main problem faced by spectropolarimetric observations of the
quiet Sun is that the signals are extremely weak (of order $10^{-3}$
of the continuum intensity or smaller). To properly detect them, long
integrations are needed. This worsens the effective spatial resolution
and the cadence of the observations, but also decreases the noise level,
which is important to cope with the different sensitivity of the
linear and circular polarization signals to weak magnetic fields in
the Zeeman regime. Due to the different sensitivity, the same
amount of noise in linear and circular polarization results in much
larger transverse fields than longitudinal fields if interpreted as a
real signal. This intrinsic bias is known to affect the determination
of the magnetic field components, particularly the field inclination,
which is usually inferred to be more horizontal than the true one
\citep[e.g.,][]{2011A&A...527A..29B}. Indeed, noise is often the main
reason for the discrepancies between analyses, hence the need to
reduce it as much as possible in the first place and then minimize its
effects with appropriate techniques.

The observations allow to constrain simulations of surface magnetism. 
Aspects that are particularly relevant for comparison with numerical
models include the spatial distribution of the fields on small
scales, the mean longitudinal flux density, the magnetic field strength and 
inclination, their stratification with height, the
temporal evolution of the fields (in particular the flux emergence
modes), and solar cycle variations (total unsigned flux, polarity imbalance, latitudinal distribution).

The main results derived from the observations are summarized below.

\begin{figure}
\begin{center}
\resizebox{0.73\hsize}{!}{\includegraphics{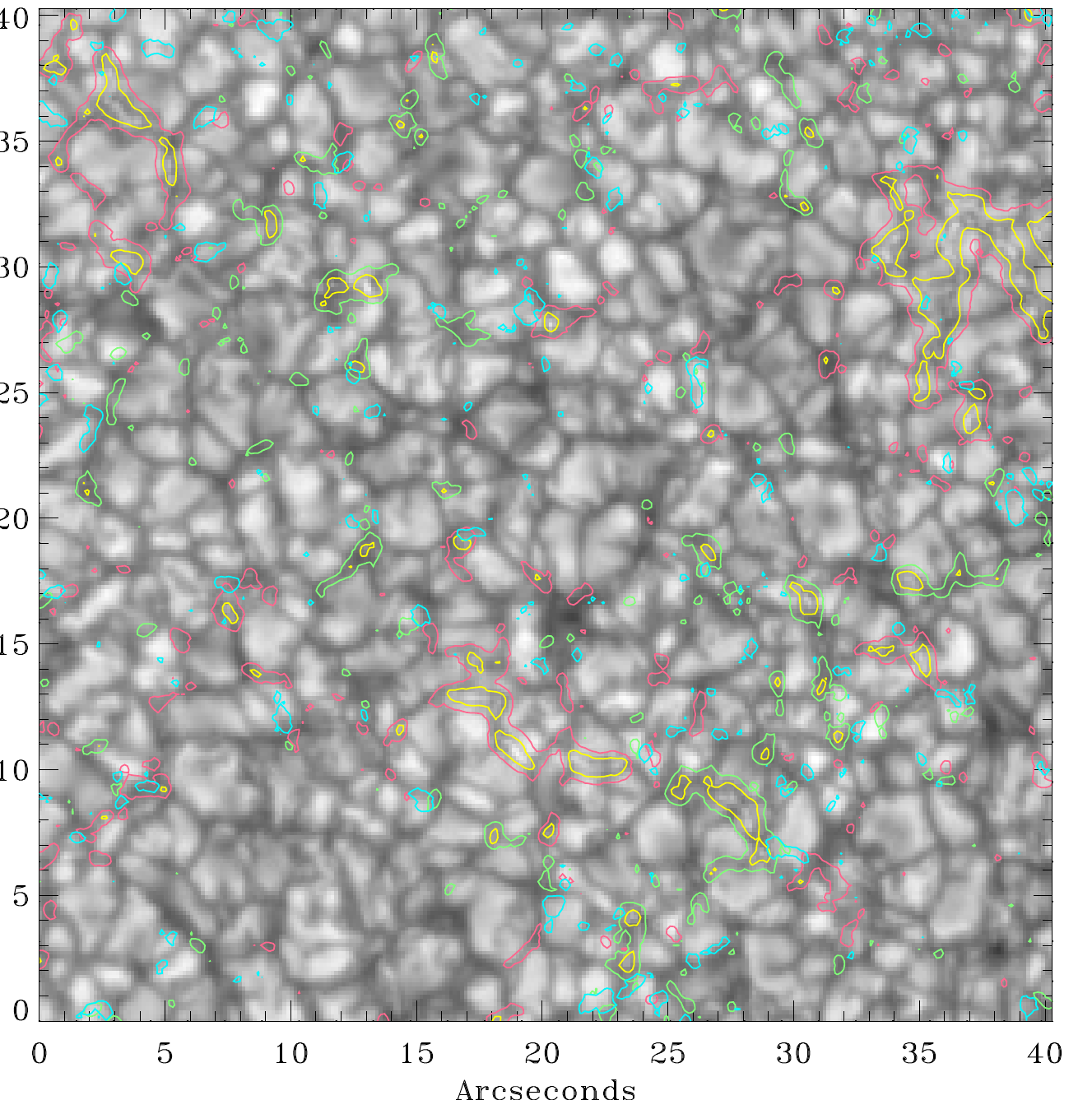}}
\caption{Spatial distribution of quiet Sun magnetic fields as observed by 
the Hinode spectropolarimeter (SP) at disk center. Red and green
contours show positive and negative apparent longitudinal flux
densities larger than 24 Mx~cm$^{-2}$ (10 times above the noise
level), while the yellow contours show strong longitudinal flux
densities of more than 100~Mx~cm$^{-2}$. Blue contours represent
apparent horizontal flux densities in excess of 122~Mx~cm$^{-2}$
(three times the corresponding noise level). The longitudinal
signals are preferentially located in intergranular lanes. The 
strong horizontal signals are spatially separated from the 
vertical signals, and are mostly seen above or at the edges
of granules. Reproduced with permission from \cite{2008ApJ...672.1237L}, copyright by AAS.}
\end{center}
\label{fig:spatial_distribution}
\end{figure}

\subsection{Properties of quiet Sun magnetic fields}
\subsubsection{Spatial distribution on small scales}

For a long time, quiet Sun magnetic fields were thought to occupy only
the intergranular space. However, with sufficient spatial resolution and
polarimetric sensitivity, granules also show clear polarization
signals. The spatial distribution of the field can be seen in
Figure~\ref{fig:spatial_distribution}. The fields in intergranular lanes tend to
be stronger, more vertical, and more concentrated than in granular
cells, which explains their higher visibility. They carry most of the
magnetic flux \citep{2009A&A...502..969B}. However, flux emergence is
observed to take place preferentially in granules in the form of
small-scale magnetic loops \citep{2007ApJ...666L.137C, 2009ApJ...700.1391M,  2020ApJ...903L..10F} and sheets
\citep{2008A&A...481L..33O, 2019A&A...622L..12F},
putting constraints on the mechanisms that bring the field to the 
surface (see Section~\ref{sect:deep_shallow} for details). Once 
at the surface, the field is dragged by the horizontal granular 
motions to the intergranular lanes, where it can be further amplified.

\begin{figure}
\begin{center}
\resizebox{0.95\hsize}{!}{\includegraphics{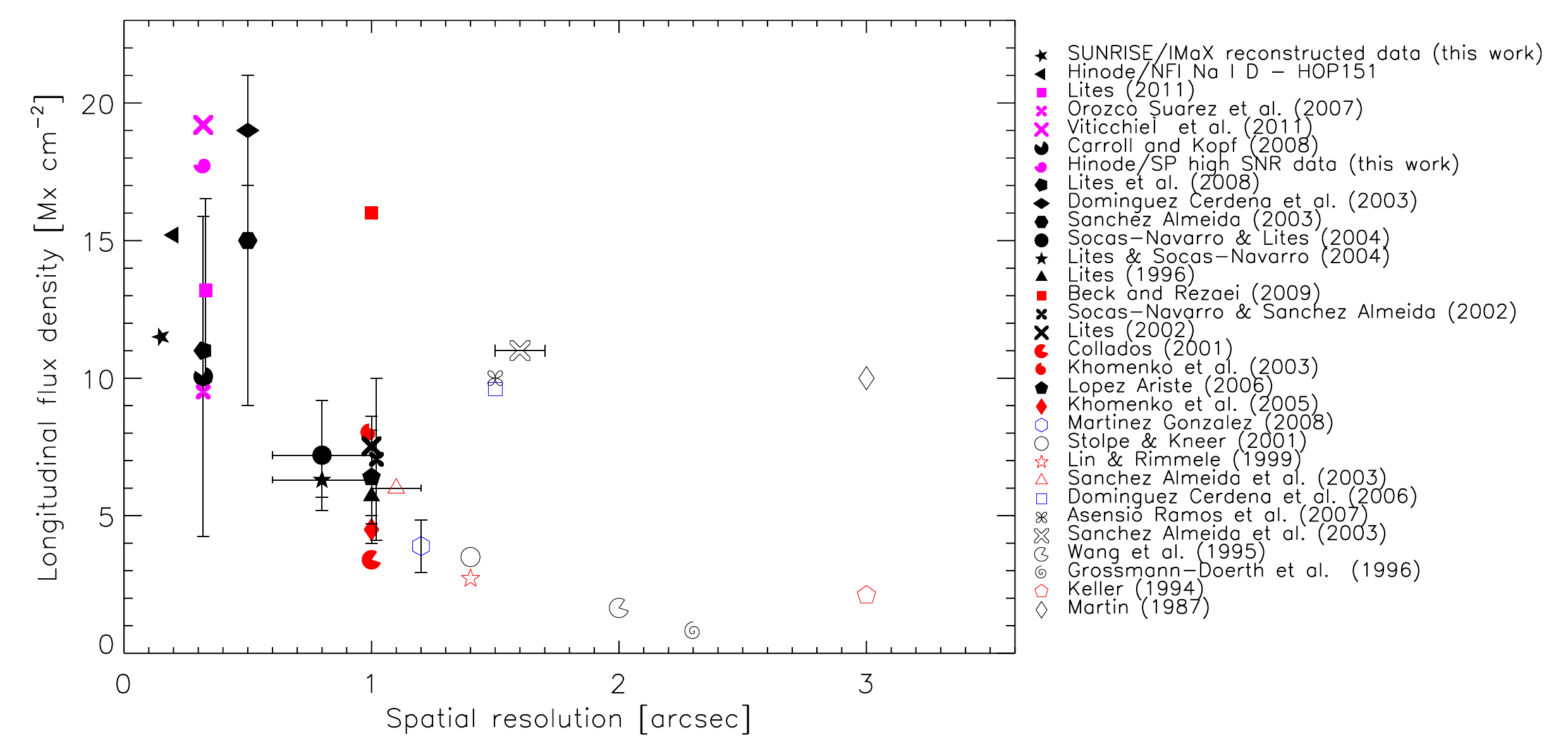}}
\end{center}
\caption{Compilation of unsigned longitudinal flux densities in the quiet Sun, 
as a function of the spatial resolution of the observations. From \cite{2019LRSP...16....1B}.}
\label{fig:fluxdensity}
\end{figure}

\subsubsection{Level of quiet Sun magnetism}

Traditionally, the magnetization of the quiet Sun has been
quantified in terms of the longitudinal flux density. This parameter
is defined as $\phi = f \, B_{\rm LOS}$, with $f$ the fraction of the
resolution element covered by the magnetic field (assumed to be
homogeneous) and $B_{\rm LOS}$ the longitudinal component of the
field. The longitudinal flux density can be derived relatively easily
from measurements of the circular polarization profile at a single
wavelength or through the inversion of full Stokes profiles, which
explains its popularity. In the weak field regime, $\phi$ is
proportional to the circular polarization signal.

Unfortunately, this parameter is very dependent on the spatial
resolution and polarimetric sensitivity of the observations. At low
spatial resolution, the magnetic filling factor $f$ tends to be small,
decreasing the flux density values. Also, the amount of Zeeman
cancellation may be significant, especially for magnetograph
observations that are not based on full spectral line profiles. This
leads to a further reduction of $\phi$. The polarimetric sensitivity,
on the other hand, affects the estimates through the noise: the larger
the noise is, the higher the mean flux density values will be, unless
provision is made to exclude pixels without clear polarization signal 
from the analysis. Therefore, high spatial resolution and high sensitivity 
are essential for reliable estimations of the flux density in the quiet
Sun.

With increasing spatial resolution, the mean unsigned longitudinal
flux density derived from Zeeman-sensitive spectral lines increases
until approximately 0.5 arcsec, where it levels off and seems to
remain constant at about $10-20$~Mx~cm$^{-2}$
(Figure~\ref{fig:fluxdensity}). The apparent longitudinal flux
densities of $7-11$~Mx~cm$^{-2}$ reported by
\cite{2010A&A...513A...1D} from Hinode/SP measurements using the
method of \cite{2008ApJ...672.1237L} are consistent with these
values. 

However, despite the apparent agreement between the estimates reported
at high spatial resolution, it is important to remember that the
longitudinal flux density only provides a lower limit to the intrinsic
longitudinal field, as it also depends on the actual filling factor of
the observations, which is unknown but certainly different from
1. This dependency makes it very difficult to compare the observed flux
densities with simulations, where the magnetic filling factor is always
unity. The magnetic field strength is less problematic and should be 
preferred for quantitative analyses, especially now that powerful 
techniques are available to retrieve it from the observations.

\subsubsection{Field strength distribution}

Stokes inversions of varying degrees of sophistication have been used
to determine the magnetic field strength, field inclination and
magnetic filling factor on a pixel by pixel basis from
spectropolarimetric observations of Zeeman sensitive lines. Stokes 
inversions allow to disentangle the actual contribution of each of
these parameters to the longitudinal magnetic flux, providing much
richer information. Thus, they are more appropriate for comparison
with simulations.

\begin{figure}
\begin{center}
\resizebox{0.99\hsize}{!}{\includegraphics{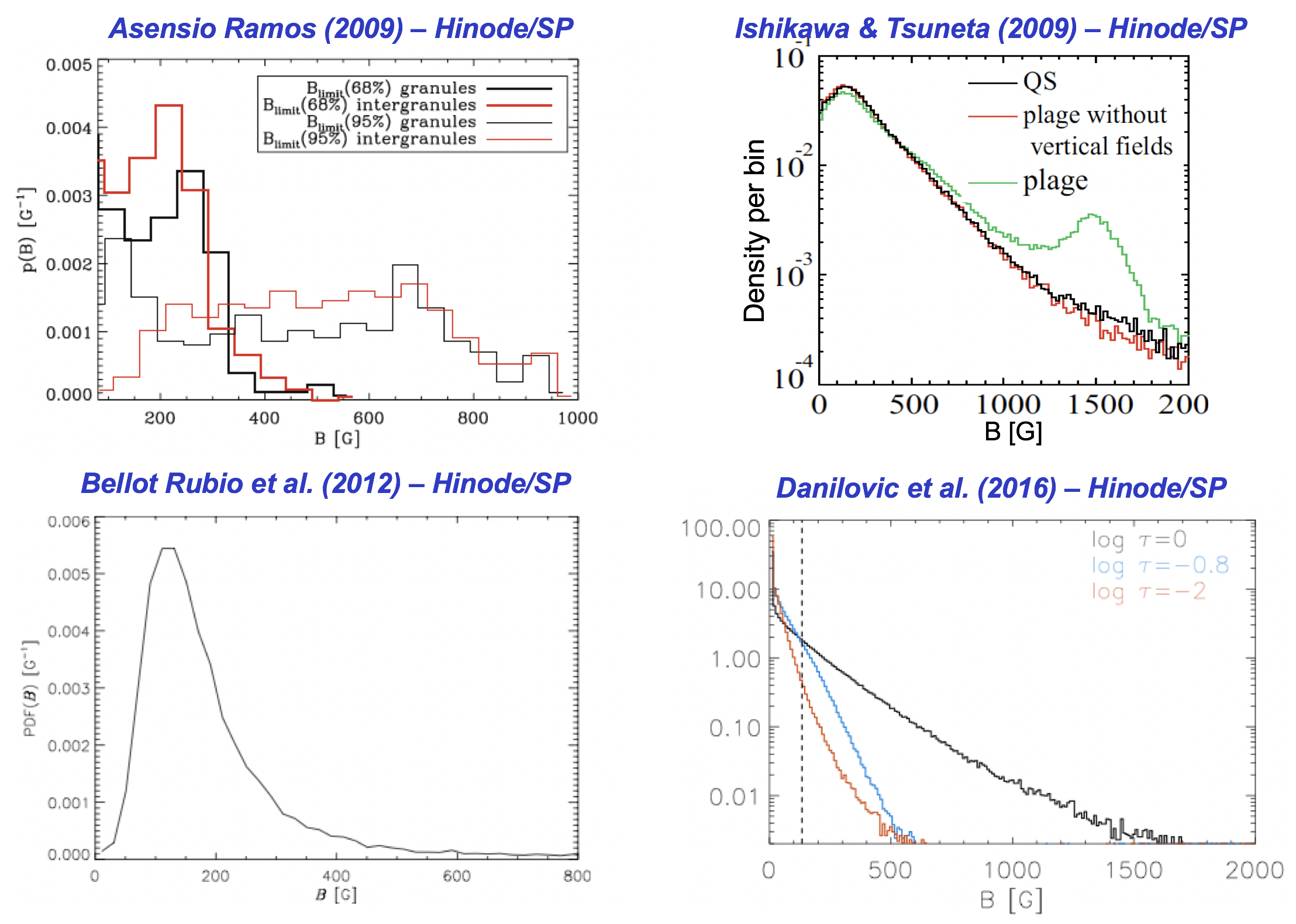}}
\end{center}
\caption{Field strength distributions in the IN determined 
from four different inversions of Hinode/SP measurements. Clockwise from top left: \cite{2009ApJ...701.1032A}, \cite{2009A&A...495..607I}, \cite{2012ApJ...757...19B}, and \cite{2016A&A...593A..93D}. Reproduced with permission, copyright by AAS (panels a, c) and by ESO (panels b, d). }
\label{fig:field_strength}
\end{figure}

According to the inversions, IN fields are weak for the most
part (Figure~\ref{fig:field_strength}). Although the details vary 
between analyses, this conclusion
seems to be robust. The field strength distribution shows a
preponderance of weak fields in the hG range and a long tail
toward kG fields. Network fields tend to be stronger than
IN fields, with a hump at 1--1.5 kG. For a summary of these 
results, see Section 4.5 of \cite{2019LRSP...16....1B}.

From the inferred field strength distribution it is possible to
compute the mean field strength over the observed field of view. The
resulting values are influenced by the noise of the Stokes profiles,
the diagnostic technique employed, and even the way the averaging is
done (in particular whether pixels with noisy signals are included or
excluded from the analysis). Also the inclusion or exclusion of the
network influences the results. But they tend to be much larger than
the average longitudinal flux densities shown in Figure~\ref{fig:fluxdensity}.
\cite{2008ApJ...672.1237L} reported a value of $\langle B\rangle = 185$~G, 
while \cite{2012ApJ...751....2O} found $\langle B\rangle = 220$~G with
$\langle \vert B_z \vert \rangle =64$~G, and \cite{2016A&A...593A..93D} derived
$\langle B\rangle = 130$~G at $\tau=1$ with $\langle \vert B_z \vert \rangle =
84$~G. These values represent upper limits to the true mean field
strengths, as in the first cases only pixels with clear signals
were considered (hence biasing the mean toward the stronger fields)
and in the last case all pixels were included (hence adding some contribution 
from photon noise in pixels with no polarization signal). In general, the 
field strength is found to decrease with height in the photosphere
\citep{2016A&A...593A..93D}. Hanle-effect inversions of molecular
lines also show a rapid drop of the field strength with height, from
95~G at $z=200$~km to 5~G at 400~km \citep{2012A&A...547A..38M}.

These results are compatible with the average field strengths
determined from spatially unresolved scattering polarization
measurements of the Sr I 460.7 nm line using the Hanle effect.
The observed center-to-limb variation of the fractional linear
polarization of the Sr I line can be reproduced by a volume-filling
magnetic field with an isotropic distribution of orientations and a
homogeneous strength of 60~G or, alternatively, an exponential
distribution of field strengths with $\langle B\rangle = 130$~G
between 200 and 400 km above $\tau=1$ \citep{2004Natur.430..326T}. 
It was concluded that most of the fields contributing to the 
Hanle depolarization of this line are located in intergranular space 
and have strengths between 2 and 300~G, well within the Hanle 
saturation regime. Fields above granules are much weaker 
and do not seem to contribute significantly to the observed 
depolarization. By including this small contribution in the
fit to the center-to-limb variation of the Sr I linear
polarization, an exponential distribution with $\langle
B\rangle = 15$~G was inferred in granular cells.

Recent results from multi-line inversions of intensity profiles
aimed to avoid the problems of noise in the Stokes polarization
spectra also confirm that granules harbor weaker fields
\citep{2021ApJ...915L..20T}. According to these inversions, the average 
field strength in granules and intergranular lanes is 16~G and 76~G,
respectively, in the optical depth range from 1 to 0.1. The average
field strength across the field of view is 46~G.

\subsubsection{Field inclination distribution}
The existence of inclined fields in the quiet Sun was known
since the mid 1990s, when small patches of Horizontal Internetwork
Fields were discovered and characterized using full Stokes
spectropolarimetric measurements at a resolution of about 1 arcsec
\citep{1996ApJ...460.1019L}. Further analyses at similar resolutions
but based on different spectral lines and inversion codes confirmed
them \citep[e.g.,][]{2003A&A...408.1115K, 2007A&A...469L..39M,
2009A&A...502..969B}.

With the significantly better spatial resolution of 0.3 arcsec
provided by the Hinode spectropolarimeter, the transverse apparent
flux density in quiet Sun areas at the disk center was found to be
about 5 times larger than the longitudinal apparent flux density,
suggesting that most of the IN fields are actually very
inclined \citep{2008ApJ...672.1237L}. Similar results were obtained
also from the ground \citep{2009A&A...502..969B}. 

\begin{figure}
\begin{center}
\resizebox{0.99\hsize}{!}{\includegraphics{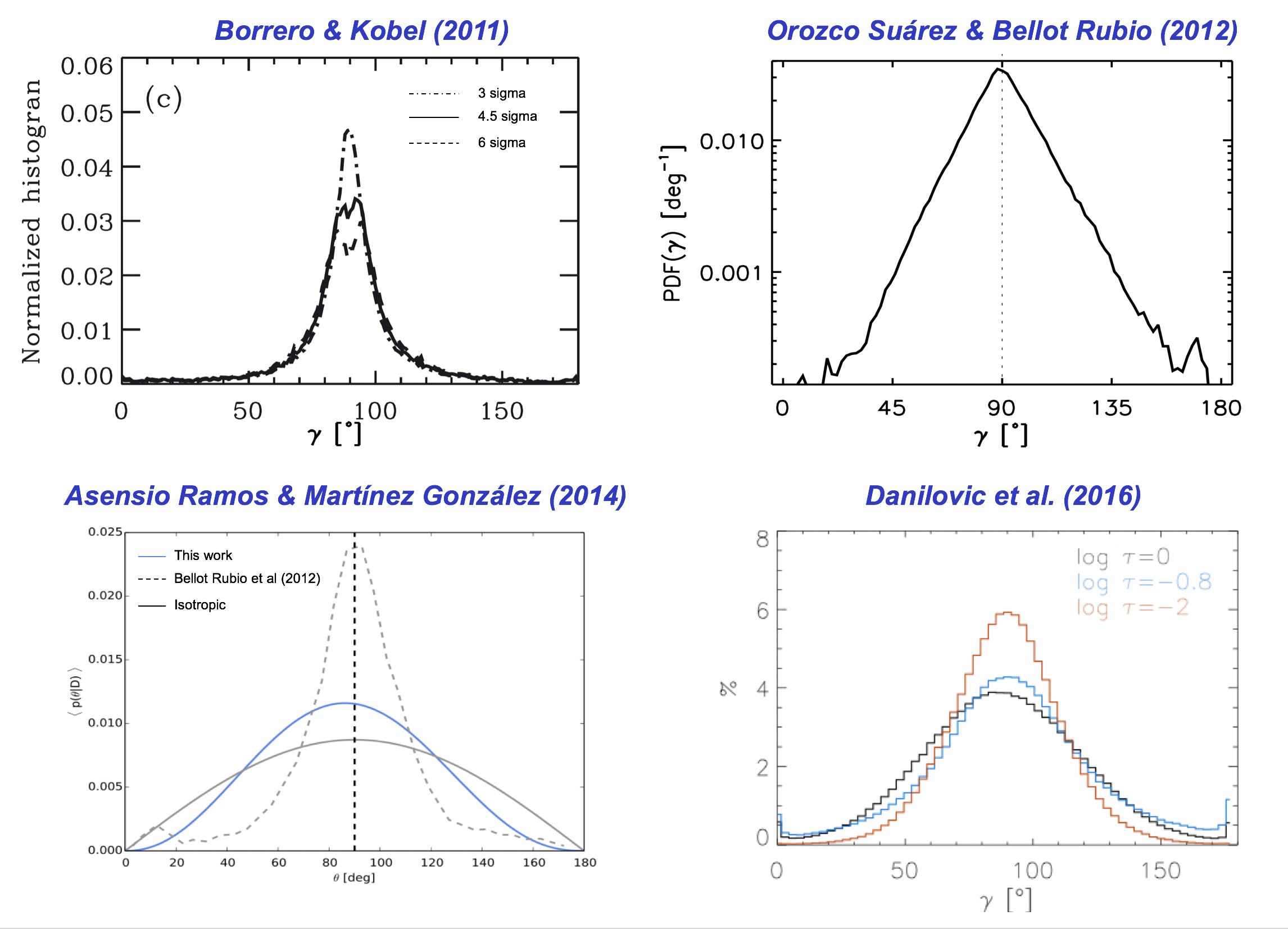}}
\end{center}
\caption{Field inclination distribution in the solar IN determined 
from four different inversions of Hinode/SP measurements at disk center. Inclinations are measured with respect to the local vertical, 
with $90^{\circ}$ corresponding to horizontal fields and 
0$^{\circ}$/180$^\circ$ representing fields pointing to/away 
from the observer. Clockwise from top left: \cite{2011A&A...527A..29B}, \cite{2012ApJ...751....2O}, \cite{2014A&A...572A..98A},  
and \cite{2016A&A...593A..93D}. Reproduced with permission, copyright by ESO (panels a, c, d) and by AAS (panel b).}
\label{fig:field_inclination}
\end{figure}

Nearly all the inversions performed to date, using different model atmospheres, 
codes and assumptions, result in a field inclination distribution at the disk 
center dominated by highly inclined fields \citep[e.g.,][]{2007ApJ...670L..61O, 2008ApJ...672.1237L, 2009A&A...502..969B, 2012ApJ...751....2O, 2014A&A...572A..98A, 2016A&A...593A..93D, 2016A&A...596A...5M}. Some examples are given in Figure~\ref{fig:field_inclination}. The field inclination distribution usually 
shows a maximum at 90$^\circ$, representing horizontal fields, and has tails 
decreasing toward 0$^\circ$ and 180$^\circ$ (vertical fields). 
While some contamination by noise cannot be ruled out \citep{2011A&A...527A..29B}, 
it is unlikely that all the inclined fields inferred in the solar IN 
are a consequence of noise in the linear polarization measurements. 
Still, the exact shape of the inclination distribution is a matter of debate, 
particularly the amplitude of the peak at 90$^\circ$, which shows significant 
differences between analyses. Solving this question requires higher sensitivity 
observations such as those to be provided by DKIST \citep{2020SoPh..295..172R} 
and EST \citep{2022A&A...666A..21Q}. 

Also, a discussion on whether the field inclination distribution is isotropic, 
quasi-isotropic, dominantly horizontal, or dominantly vertical is ongoing 
\citep[see][]{2019LRSP...16....1B}. This is an important question 
that can shed light on the origin of the fields or the way they appear 
on the solar surface. The answer may be  different at different atmospheric 
heights. For example, an isotropic distribution is observed in simulations of 
small-scale dynamo action in the low photosphere, even if convection is 
anisotropic \citep{Rempel:2014:SSD}. On the other hand, a predominantly 
horizontal distribution may be the consequence of the field being organized 
in magnetic  loops on granular scales, as suggested by, e.g., \cite{2008ApJ...672.1237L} 
and \cite{2019LRSP...16....1B}. Investigating this issue requires 
observations outside of the disk center. If the field is isotropic,
the inclination distribution should not change with the heliocentric
angle. Unfortunately, studies of the center-to-limb behavior of the
inclination distribution are very scarce \citep[e.g.,][]{2012ApJ...746..182O}. 
The efforts have rather focused on determining the variation of the 
circular and linear polarization amplitudes, since they are not biased 
by photon noise. The results of these analyses are still
controversial, but there seems to be a variation of the weakest
polarization signals with heliocentric angle which would not be
compatible with an isotropic distribution of field orientations
\citep{2008ApJ...672.1237L, 2013A&A...550A..98B, 2014PASJ...66S...4L, 
2017ApJ...835...14L}.

Finally, the inclination of IN fields appears to vary with
height, becoming more horizontal in the mid photosphere
\citep{2016A&A...593A..93D}. This may just be the result of the 
existence of small-scale loop-like structures straddling a few
granules all over the solar surface, as the loop tops naturally have
more horizontal fields and are located higher in the atmosphere than
the footpoints. The MHD simulations of 
\cite{2008ApJ...680L..85S} show
a predominance of inclined fields at a height of 500~km, where the
ratio of horizontal to vertical field components is 2 to 5.6, in
accordance with \cite{2008ApJ...672.1237L}. In the simulations, the
horizontal field strength reaches a maximum in the upper photosphere 
because overshooting convective motions expel the horizontal field 
upwards to layers where vertical flows are no longer present, allowing 
the field to accumulate there. This purely dynamic effect is also 
observed in SSD simulations, resulting in fields that are more 
horizontal in the upper photosphere and ratios of horizontal to 
vertical field that are consistent with the
observations \citep[][see Section~\ref{sect:anisotropy}]{Schuessler:Voegler:2008:bhorz, Rempel:2014:SSD}. 

\subsection{Flux emergence in the quiet Sun}
\begin{figure}[p]
\begin{center}
\resizebox{0.9\hsize}{!}{\includegraphics{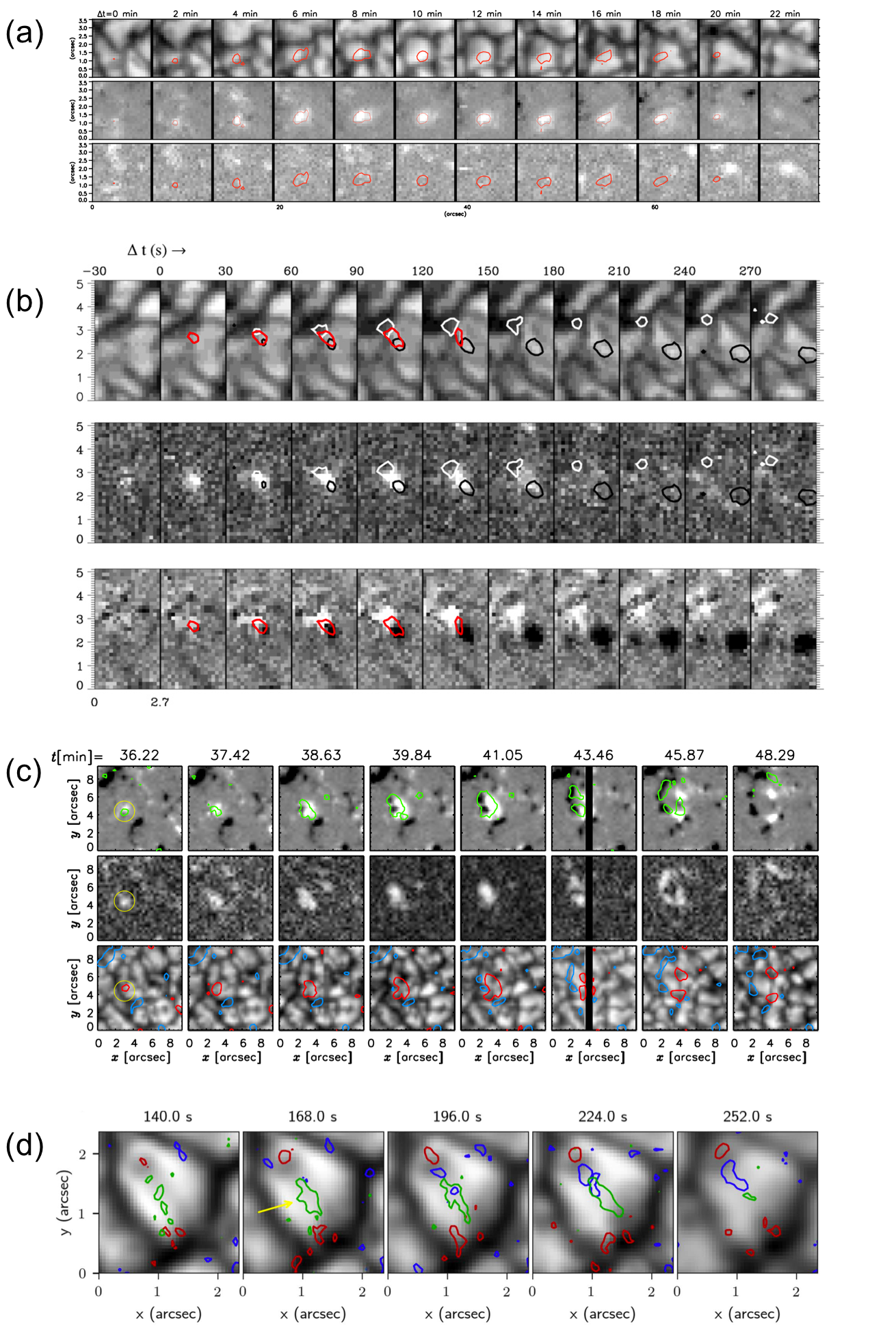}}
\end{center}
\caption{Examples of magnetic flux appearance in the quiet Sun. (a) Unipolar features. (b) Bipolar magnetic loops. (c) Sheets of horizontal fields covering a granule. (d) Horizontal fields in granular lanes. The different rows show continuum intensity, circular and linear polarization in (a), continuum intensity, linear and circular polarization in (b), circular and linear polarization and continuum intensity in (c), and continuum intensity in (d). Circular polarization patches are indicated with red contours in (a), white and black contours in (b), and red and blue contours in (c) and (d). Linear polarization patches are outlined with red contours in (b) and green contours in (c) and (d). Adapted from \cite{2008A&A...481L..33O}, \cite{2009ApJ...700.1391M}, \cite{2019A&A...622L..12F}, and \cite{2020ApJ...903L..10F}. Reproduced with permission, copyright by ESO (panels a, c) and by AAS (panels b, d).}
\label{fig:flux_emergence}
\end{figure}

Magnetic flux emergence is an ubiquitous process in the quiet Sun. It
happens on a wide range of spatial scales (from mesogranular to
granular and subgranular scales) and on short timescales, but long
duration observations are needed to characterize its statistical
properties. Such observations are difficult to obtain, as they require
stable conditions for hours. IN magnetic fields are observed
to appear on the solar surface in two flavors: as individual bipolar 
features or clusters of mixed-polarity elements, and as unipolar patches. 
The latter are features of given polarity without any clear associate
opposite-polarity element in the surroundings 
(Figure~\ref{fig:flux_emergence}a). It has recently been
shown that about 55\% of the total IN flux appears in
bipolar form, while the rest is unipolar \citep{2022ApJ...925..188G}. 
The physical properties of these two populations turn out to be 
different, which suggests different origins.

Studying the modes of appearance of quiet Sun fields before they
interact with photospheric convective flows is key to understanding
their nature through comparisons with numerical simulations. 
Small-scale bipolar emergence in seen to occur in the form of 
magnetic $\Omega$-loops \citep{2007ApJ...666L.137C, 2009ApJ...700.1391M} 
and magnetic sheets  \citep{2019A&A...622L..12F}. Magnetic 
loops emerge into the photosphere above or at the edges of granules, 
showing linear polarization in between two-opposite circular
polarization patches (Figure~\ref{fig:flux_emergence}b). These 
signals represent the horizontal field 
of the loop top and the vertical field of the loop legs,
respectively. The linear polarization signals are conspicuous features
that show up prominently in high-sensitivity spectropolarimetric
observations \citep{2010ApJ...723L.149D, 2010ApJ...718L.171I,
2011ApJ...735...74I, 2012ApJ...755..175M, 2018SoPh..293..123K,
2021ApJ...911...41G}. It has been suggested that the magnetic topology
of these small-scale loops may explain the field strength and field
inclination distributions observed in the quiet Sun IN \citep{2019LRSP...16....1B}.

Another type of bipolar flux emergence in the quiet Sun involves
large sheets of horizontal fields covering a full granule
\citep[][see Figure~\ref{fig:flux_emergence}c]{2019A&A...622L..12F}, 
which has been identified also in simulations 
\citep{2018ApJ...859L..26M}. The sheet fragments as it expands to the
granular edges, leaving only the footpoints that can be observed as
opposite-polarity patches in circular polarization maps. This form of
flux emergence may explain the small clusters of mixed-polarity
elements observed in longitudinal magnetograms such as those analyzed by
\cite{2022ApJ...925..188G}, but an unambiguous confirmation is not 
possible until the appearance sites of the cluster members are
determined. 

Finally, horizontal magnetic fields flanked by vertical fields have been
observed to emerge also in granular lanes produced by horizontal
vortex tubes \citep{2020ApJ...903L..10F}. An example is shown in Figure~\ref{fig:flux_emergence}d. The magnetic field has a
loop-like structure with strengths of several hundreds Gauss and is
detected at the late stage of the granular lane development. The vortex
tube grabs preexisting horizontal fields in the adjacent intergranular
lanes, located at or below the surface, and takes them to the granular
interior, where they are transported to the surface and eventually
back to the intergranular space by the granular upflows. In this way,
vortex tubes provide a mechanism for the local recirculation of 
magnetic field required by the small-scale dynamo to operate 
on the solar surface (see Section~\ref{sect:deep_shallow}). 

While bipolar flux emergence should be considered the dominant 
form of flux appearance in the quiet Sun, unipolar appearances 
still provide a considerable fraction of the magnetic flux 
present on the solar surface. However, they pose an important 
problem, as the opposite polarity that must be associated 
with every unipolar feature seems to be missing. Clearly, 
this is a detection problem. Very likely, the signals are there but
cannot be seen due to insufficient polarimetric sensitivity. Indeed,
it has been suggested that unipolar features do not bring new magnetic
fields to the solar surface, but are the result of very weak
background flux that is hidden in the noise until some mechanism
concentrates it, becoming visible above the detection threshold
\citep{2008ApJ...674..520L, 2022ApJ...925..188G}. The exact mechanism 
is presently unknown, but it might be related to convectively driven
sinks at mesogranular vertices \citep{2017ApJS..229...14R}. Also the nature
of the background flux is unknown, in particular whether it would be  
produced by small-scale dynamo action or by the large-scale dynamo. 

Upon appearance on the solar surface, magnetic fields interact with
the local granular motions and are dragged by supergranular flows
towards the edges of the supergranular cells \citep[e.g.,][]{2014ApJ...797...49G}. 
Bipolar features can be followed for some time until the footpoints 
cancel or merge with other magnetic features, losing their identity. 
Both the magnetic topology and the evolution of these features carry 
information on the origin of the fields, and are therefore important 
parameters to be compared with numerical simulations. A specific open 
question is the role of IN fields in the formation of the quiet 
network outlining the boundaries of supergranular cells. This will 
be briefly discussed in the following section.

\subsection{Contribution of the IN to the quiet Sun network} 
\label{sect:network_formation}

The quiet Sun network is believed to be formed by active region decay, 
ephemeral regions and IN fields. It shows a variation with the 
solar cycle, which reflects its connection with the large-scale 
dynamo (see Section~\ref{sec:QS_cycle_variation}). However, 
the contribution of the various components to the network 
flux is still under discussion. 

It has been shown using Hinode/SOT data that about 40\% of the total
IN flux eventually ends up in the quiet network \citep{2014ApJ...797...49G}.  
According to those observations, the IN transfers magnetic flux to the 
network at a rate of $1.5 \times 10^{24}$~Mx~day$^{-1}$ over the 
entire solar surface. This means that the IN supplies as much flux as 
is present in the network in only 9-13 hours, and could maintain it. 
 
The results of \cite{2014ApJ...797...49G} suggest that the  
IN is an important source of flux for the network, in agreement 
with the increasing evidence of a surface SSD contributing to the 
network field \citep{Rempel:2014:SSD}. On short-time scales, the 
IN flux transferred to the network may provide the seed 
for further amplification of the field up to kG values at the site of 
converging mesogranular and supergranular flows \citep{2017ApJS..229...14R,
2018A&A...610A..84R}, explaining the larger abundance of strong,
long-lived flux concentrations in the network compared with the
IN. This process would be consistent with the first mechanism of network 
formation described in Section~\ref{sect:network}. 
 
To verify or refute this idea, an observational investigation of the
appearance and evolution of magnetic flux in the network must be
carried out, considering also the adjacent IN. Particularly
important is the site of appearance of new flux within the granulation
pattern, as well as the interaction between network and IN 
fields, with a view to determine how the flux is eventually deposited
in the network. Such an analysis has never been performed at the
required spatial resolution and sensitivity.

\subsection{Solar cycle variations of quiet Sun fields}
\label{sec:QS_cycle_variation}

Despite being plagued with difficulties, the search for possible
variations of the quiet Sun magnetism with the solar cycle has been
pursued vigorously in the last decades. This is because the detection
of temporal and/or latitudinal variations would link the quiet Sun
magnetism with the large-scale dynamo responsible for the solar magnetic
cycle. The main challenge is the need of very stable, homogeneous 
observations over periods of time spanning years. Few
instruments are capable of providing such observations. Space-borne
measurements are preferred, but also ground-based Hanle effect 
observations have been used to that end.

As summarized in Section 3.2 of \cite{2019PASJ...71R...1H}, evidence
for temporal variations of the polarization signals produced by quiet
Sun fields is very marginal, if present at all. Most studies do not
find significant changes or they are within the statistical
uncertainties of the observations.

An analysis of 5.5 years of Hinode/SP data taken in quiet regions at
the disk center revealed no measurable variation of either the
magnetic flux or size distribution of IN patches with time
\citep{2013A&A...555A..33B}. Similar Hinode/SP observations were used
to derive the transverse and longitudinal magnetic flux densities of
very weak IN regions at different positions on the solar
disk and study their variation from 2008 to 2015
\citep{2015ApJ...807...70J}. No change in the flux density or the ratio
of transverse to longitudinal flux was found. Following a different
approach, synoptic Hinode/SP observations taken at various positions
along the central meridian were used by \cite{2014PASJ...66S...4L} to
investigate the long-term evolution of the magnetic flux density in 
quiet IN regions. The transverse and unsigned longitudinal 
fluxes were found to be independent of the solar cycle at all 
solar latitudes, while the signed longitudinal fluxes (i.e., the 
polarity imbalance) showed clear changes in the polar regions and 
some hints of variation in the activity belt from 20 to 60 degrees latitude.

Full-disk observations in the near-infrared Fe I line at 1564.8~nm did
not show notable changes in the properties of the polarization signals
from IN regions for much of solar cycle 24
\citep{2020ApJ...904...63H}.

\cite{2021A&A...651A..21F} performed a Fourier analysis of the spatial 
fluctuations of the longitudinal flux density in small $10'' \times
10''$ IN regions along the central meridian, using synoptic
data from the Hinode/SP between 2008 and 2016. On scales smaller than
$0.5''$ they did not find significant variations of the magnetic
fluctuations with the solar cycle at any latitude. On granular scales,
up to about $2.5''$, the power of the spatial fluctuations did not
show variations at low and mid latitudes either, but a decrease was
observed at high latitudes during solar maximum. The lack of changes
on scales smaller than $0.5''$ indicates the presence of a
time-independent magnetic field in the IN. However, the
variation detected on larger scales at high latitude suggests that 
also the large-scale dynamo contributes to the magnetism of the IN,
although not homogeneously over the solar surface.

Using 12 years of SDO/HMI data, the rms longitudinal flux density in 
quiet $1^{\circ} \times 1^{\circ}$ IN regions at the central
meridian was found to be nearly constant over time, with an
average value of 6~Mx~cm$^{-2}$ \citep{2022A&A...665A.141K}. This 
was interpreted to reflect a real independence of the quiet IN 
magnetism on the solar activity cycle, or the inability of HMI 
to detect changes due to insufficient sensitivity. By contrast, 
the rms flux density in $15^{\circ}$ windows did show a 
statistically significant correlation with the solar cycle, 
with the maximum of the curve lagging the sunspot number cycle 
by about half a year. The difference is that these $15^{\circ}$ 
windows contain both network and IN fields, whereas the 
$1^{\circ}$ windows avoid the network fields. This suggests 
that the quiet network is indeed affected by the large-scale solar
dynamo, although on supergranular timescales its evolution seems 
to be determined by interactions with IN fields, presumably 
reflecting a contribution from the small-scale dynamo (see Section~\ref{sect:network_formation}). \cite{2019RAA....19...69J} also
found a variation of the network with the solar cycle using full-disk 
SDO/HMI magnetograms, but in this case the quiet network flux 
showed an anti-correlation with the sunspot number, due primarily 
to a reduction of the network area (the magnetic flux density of 
quiet network patches was observed to increase by about 6\% at 
solar maximum). Further analyses are needed to clarify the exact
variation of the quiet network fields with the sunspot cycle and 
the contribution of the small-scale dynamo to its formation and maintenance. 

Following a different strategy, the scattering polarization 
measurements of \cite{2010A&A...524A..37K} did not show
significant changes in the amount of Hanle depolarization of selected
$C_2$ molecular lines over two years spanning the minimum phase of 
solar cycle 23. The observed lines are formed almost exclusively in
granules, so they sample only the weakest fields of the quiet Sun.
The latest analysis, covering almost a full solar cycle, still shows 
no clear changes in the Hanle depolarization with time 
\citep{2019ASPC..526..283R}. It is important to continue this type 
of synoptic Hanle programs and possibly extend
them to the photospheric Sr~I~460.7~nm line (which is formed also in
the intergranular lanes), as they provide an independent way to
examine the cycle-dependence of the quiet Sun magnetism.

All these results point to no or little variations of the weak IN
fields with the solar cycle, which supports the view that they are
generated by an SSD and not by an LSD cascading down to smaller
spatial scales. By contrast, the quiet network seems to show a variation 
with the solar cycle, indicating some contribution from the LSD
in addition to a possible one from the SSD, but its amplitude and 
phase are not well determined yet. 

\section{Basic theoretical concepts}\label{sect:theory}
We start our discussion about the theoretical and numerical studies of SSD with a short summary of the basic theoretical concepts. We then consider more idealized numerical models, used for studying the basic properties and parameter dependencies of SSD and comparisons with theory, and finally move on to surface convection simulations, which are calibrated and can be compared against observational data.

A SSD refers to the sustained and rapid amplification of magnetic field fluctuations at spatial scales smaller than the forcing scale in a plasma system.
In the case of turbulent convection, the forcing scale is
the scale at which the kinetic energy spectrum peaks, which is the scale of convective cells, that is believed to vary strongly as function of depth: according to the mixing-length theory, the convective cells are small and turn over fast near the surface, and get progressively larger and slower as function of depth \citep{Vitense53}.
In the solar convection zone,
the plasma is turbulent enough to rapidly amplify magnetic fluctuations 
at or smaller than the scale of convective cells. 
The environment required for this dynamo instability to operate is such that
the influence of rotation is weak and the flow is largely non-helical. This is in contrast to a large-scale dynamo (LSD) where the amplification of fields occurs at scales larger than the forcing scale, which is a consequence of symmetry breaking (due to helicity, inhomogeneities, anisotropies, etc.) 
at small scales, facilitating an inverse cascade of magnetic energy from small to large scales, further assisted by large-scale non-uniformities in the velocity field
\citep[see, e.g.][]{Ch20}. 
This would suggest that the LSD would preferentially occur in the deeper layers of the CZ, where rotational influence on convection is strong, while SSD would operate nearly unimpeded in the surface layers, where the turbulent part of LSD would have only little chances of existing. 
As already hinted towards from observations, the situation is likely to be much more complex, with these two dynamo instabilities being excited together over a large fraction of the CZ, and non-linearly influencing each other.

Let us start our brief theoretical discussion by introducing the most important dimensionless control parameters of SSD-active systems. 
In the following,
we use $\eta$ for magnetic diffusivity, $\nu$ for molecular viscosity and $U$ as the typical velocity at the largest scale, $L$, of the inertial range. The latter also enters discussion of power spectra in terms of the scale of forcing/energy injections as $k_i \sim 1/L$. Then, the fluid Reynolds number $\Rey = UL/\nu$ and the magnetic Reynolds number $\Rm = UL/\eta$. 
The magnetic Prandtl number is defined as the ratio of the two, namely $\Pm = \Rm/\Rey$, and can hence also be written as $\Pm=\nu/\eta$.
The Reynolds numbers in the solar plasma are both large, but the magnetic one is estimated to be a few orders of magnitude smaller than $\Rey$, hence leading to $\Pm$ values in the range of $10^{-6} ... 10^{-4}$ \cite[see, e.g.,][]{BS05,SchSre2020}.

\begin{figure}
\begin{center}
\resizebox{0.75\hsize}{!}{\includegraphics{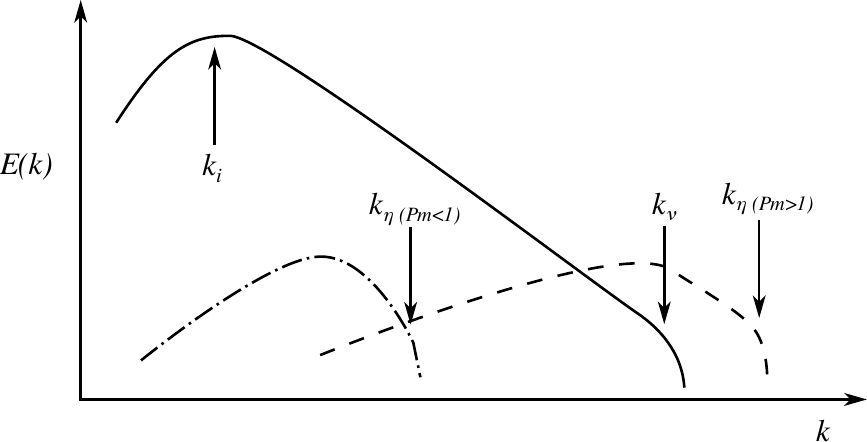}}
\end{center}
\caption{Schematic of power spectrum for magnetic and kinetic energy in the low and high Pm regime. The solid line illustrates the kinetic energy for a turbulent plasma with 
energy injection at spatial wave number $k_i$ and dissipation at $k_\nu$. The dot-dashed (dashed) line show the magnetic energy power spectra in the kinematic growth phase with 
dissipation at scales $k_{\eta (\Pm<1)}$ ($k_{\eta (\Pm>1)}$) for the Pm$>1$ (Pm$<1$) case.}
\label{fig:ps_scheme}
\end{figure}

\cite{Bat1950} discussed the possibility of magnetic field amplification in a turbulent flow by drawing an analogy between the time evolution of vorticity in incompressible turbulence and the induction equation for the magnetic field. 
His model predicted no SSD for $\Pm<1$ plasmas, however.
The first rigorous mathematical treatment was performed by \cite{K68}, where it was shown that for the simple case of Gaussian zero-mean, homogeneous and isotropic velocity field that is $\delta-$correlated in time (Kraichnan ensemble \citep{Kra1968}), the evolution of magnetic energy (or equivalently, the magnetic correlation tensor) can be expressed in a form similar to the Schr\"odinger equation with an effective "mass" and a "potential" that depends on the velocity correlation tensor. A description of this tensor then completes the system. The simplest assumption is to take the scale dependence of velocity fluctuations as $\avg{\delta u(l)} \sim l^\alpha$, where $\alpha$ can range from 0 to 1, corresponding to a "rough" and "smooth" velocity field, respectively. The bound-state solutions of the equation, then, describe exponentially growing modes. When $\eta$ is non-zero, this potential becomes repulsive at both the smallest and largest scales, allowing dynamo action to take place only if there is sufficient scale separation between the integral and the dissipative scales (see, e.g., section 3.2 and figure 4 of \cite{Vin2001} for an illustration). In other words, there exists a critical magnetic Reynolds number $\Rmc$ corresponding to this scale separation and dynamo action is possible only when $\Rm > \Rmc$.
This quantity depends on $\Rey$ through $\Pm$, the latter of which is independent of scales and flow properties.

According to Kazantsev's model, the amplification occurs at the timescale of the turbulent eddy turnover time, which is very short in comparison to the timescales required for the amplification of the large-scale field. In the kinematic growth phase, when the magnetic fluctuations are still weak, the structures generated have the thickness around the resistive scale, but are curved up to the scale of the turbulent eddies. Hence, the peak in the magnetic power spectrum is at the resistive scale, but all scales grow with the same growth rate. At scales larger than the resistive scale, a positive power law of $k \propto 3/2$ is predicted \citep{K68}, hence called the Kazantsev spectrum, while at smaller scales, the spectrum can fall off very steeply, following the so-called Macdonald function \citep[for details, see e.g.][]{BS05,Rinconrev19}.

In low-$\Pm$ plasmas, as the solar one,
the resistivity, $\eta$, is much larger than the fluid viscosity, $\nu$, meaning that the dissipation of the fluid motions occurs at scales much smaller than the scales at which magnetic fields dissipate (for an illustration of how the spatial 
power spectrum
of magnetic and kinetic energy look like for low and high $\Pm$ cases, see Fig.~\ref{fig:ps_scheme}). In effect, the turbulent eddies can dissipate into much smaller-scale structures than the magnetic structures, due to which the magnetic fluctuations must grow within the inertial range of the turbulent spectrum. The smoother magnetic structure, therefore, sees the turbulent eddies as a rough field around it (comparable to the "rough" flow in the Kazantsev picture); these circumstances are to be contrasted with high $\Pm$ fluids, where a smooth velocity field would be acting on smaller magnetic structures. The amplification of the magnetic fluctuation is more challenging in the former case of a rough velocity field, and hence the critical Reynolds number for dynamo action, $\Rmc$, is elevated. 
At the incompressible limit or near it (weak compressibility),
$\Rmc$ ranges between 30--60 for high-$\Pm$ plasma \citep[e.g.,][]{BS05}, 
while for low $\Pm$ values of around 400 have been analytically computed \citep[e.g.,][]{KR12}.

How the SSD non-linearly saturates remains an open problem. Extending the analytical work on the Kazantsev model, it has been proposed that the SSD can grow magnetic fluctuations at the resistive scale up to and exceeding the equipartition with kinetic energy of turbulence, but that the generated field would be concentrated into resistive-scale ropes, hence not being volume filling, and therefore being energetically rather insignificant \citep[e.g.][]{Kandu98}. The non-linear regime poses a formidable problem for analytical studies, but numerical studies can be attempted. 

\section{Numerical models}\label{sect:models}

Reaching the extreme-scale $\Pm$s of the solar and stellar convection zones is impossible currently and will most likely remain so in the future, at least for explicit-diffusion codes.
The simulations conducted with these codes are also referred to as direct numerical simulations (DNS), although they are not quite fulfilling this definition in a strict sense, as orders of magnitude elevated diffusivities are used than in the real object; hence, hereafter we refer to these type of models as DNS-like.

An alternative to explicit diffusivity schemes is the usage of implicit large-eddy simulations (hereafter ILES), where the diffusive terms are replaced with numerical counterparts, providing diffusion only close to the grid scale, while leaving well-resolved scales unaffected. This has the advantage to maximize the Reynolds numbers in the flow. However, the actual values of the dimensionless control parameters then become ill-defined. There are various incarnations of these techniques, ranging from hyperviscous operators 
\cite[see, e.g.,][]{SN98}
to slope-limited 
diffusion schemes \cite[see, e.g., ][]{Rempel:2014:SSD}, to mention just a couple \cite[for a more thorough review, see][]{Miesch_LES15}.

Third type of numerical schemes are the so-called explicit large-eddy simulations (hereafter ELES), where the solved equations are filtered at a spatial scale larger than the grid scale, and (ideally) physically motivated subgrid-scale (SGS) models are used to describe the terms describing the scales left out by the filtering procedure. The best known example is the Smagorinsky scheme \cite{Smagorinsky63}, that develops a concept of turbulent eddy viscosity as an SGS model. More involved closures for the magnetized case, where more terms than the viscosity itself need to be described by the SGS model, have also been developed \cite[see, e.g., ][]{Greteetal17}, although their usage in convection modeling is still limited. In the most complicated cases the ELES should also account for the influence of small-scale turbulence on large-scale dynamics, such as the inductive action of helical small-scale turbulence on large scales, dubbed the $\alpha$ effect, usually referred to as backscatter in LES terminology. 

ILES schemes have been immensely successful for
solar surface simulations with radiation transport
\citep[e.g.][]{SN98,Rempel:2014:SSD}, as will be described later on in this chapter.
However, concrete indications that ILES schemes might not be sufficient and
appropriate for convective systems with LSD dynamo and SSD together come from comparisons like that of
DNS-type simulations of \citep{KKOWB17} and ILES-type ones of
\citep{HRY16} showing markedly different behaviour, (also to be discussed later in this chapter).
Nevertheless, numerical models give valuable insights to the regimes, where analytical approaches completely fail, such as studying the saturation mechanisms, and the interactions between SSD and LSD.

\subsection{The quest for finding and studying low-$\Pm$ small-scale dynamos}\label{subsect:Pm}

Numerical studies of SSD in solar and stellar contexts are challenging due to the low $\Pm$ of these environments. As discussed earlier, the critical $\Rm$ required for the onset of SSD mechanism is expected to be in the order of hundreds, already requiring high resolution. But there's also the added challenge of $\Rey$ required being orders of magnitude larger than $\Rm$ to reach the physically relevant low-$\Pm$ region. Hence, limited numerical studies exist with only moderately small $\Pm$, with the minimal achieved values currently being approximately 0.003 by \cite{Warnecke22}). Only a fraction of studies pushing to the low-$\Pm$ regime include convection as the driver of the background turbulence, while the majority use some sort of idealized forcing function. It has been argued, however, that the SSD properties are only weakly dependent on the type of the forcing used \citep[see, e.g.][]{Moll2011}. 
With the current computational resources we are, however, finally at the limit of being able to answer the first imminent question of whether the analytic predictions of $\Rmc$ are correct, or if the threshold is lower/even higher than expected. 

\subsubsection{Kinematic phase}

While numerical evidence for SSD at $\Pm=1$ in incompressible, homogeneous, isotropic, and non-helical setting was obtained several decades ago by \cite{Meneguzzi1981}, and the literature of high-Pm number SSD in varying 
setups is abundant \citep[see, e.g., the reviews by][]{BS05,Rinconrev19}, the first numerical evidence for low-$\Pm$ dynamos dates back to only 15 years \cite{ISCM07}. This initial evidence was restricted to simplified, forced, setups with hyperdiffusivity. The encouraging result from these studies is that the $\Rmc$ does not continue increasing with $\Pm$, but appears to have a maximum \cite{ISCM07}.
A more recent DNS-like study by \cite{Warnecke22} shows that after plateauing, $\Rmc$ even starts decreasing again,
approaching values of the order of a hundred.
This non-monotonic behavior appears now to be firmly associated with the magnetic energy peak falling into the 
so-called bottleneck range of the kinetic spectrum \cite[e.g.][]{Brandenburg:2011:SSD_low_Pm,Warnecke22}. The bottleneck refers to both experimentally \citep[reported 
in several papers since][]{SJ93} and numerically \cite[reported since][]{DHYB03} confirmed inefficiency of the kinetic energy cascade at wavenumbers somewhat smaller than the viscous scale.
This is seen most clearly by plotting the kinetic energy power spectrum (as in fig. \ref{fig:ps_scheme}) compensated by the Kolmogorov scaling, i.e., $E(k)k^{5/3}$. For ideal Kolmogorov turbulence, it should result in a flat line in the inertial scale, followed by a drop-off around the viscous scale. However experiments and simulations 
show a "hump" that forms near the viscous scale before the drop-off occurs. This hump, especially when its lower wavenumber side,
where the deviation from Kolmogorov scale is positive,
overlaps with the energy-carrying scale wavenumber of the magnetic fluctuations, has been attributed to decreased SSD efficiency \citep{Warnecke22}.

In the kinematic regime, one is additionally interested in determining the growth rate and its dependence on the key system parameters, and the evolution of the energy spectrum of the magnetic fluctuations, for which links to the Kazantzev theory can be established. Such links can be obtained by mapping an appropriate turbulence model, such as a one from Kolmogorov theory, to the Kazantsev theory. For example, the predicted growth rate by Kazantsev theory, $\gamma \propto u_{\rm d}/l_{\rm d}$, can be estimated by replacing estimates of dissipation scale velocity, $u_{\rm d}$, and length scale, $l_{\rm d}$, from Kolmogorov theory. 
Similarly, estimating $\Rmc$ for different $\Pm$ has been based on using the roughness of the turbulent flow at different scales as a mapping \citep[see, e.g., ][]{BC04}. Also, the effects of compressibility have been taken into account 
by assuming a linear relation between the transverse and the longitudinal component of the velocity correlation tensor for the two extreme cases of divergence-free (Kolmogorov turbulence) and irrotational (Burgers turbulence) flow and assuming the same form for the longitudinal component \citep[see, e.g., ][]{Schober2012}. 
For solar-like weakly compressible flows all (semi-)analytical models indicate that a kinematic dynamo exists for the case of low $\Pm$, and differ only in the details of the estimated $\Rmc$ values for different flow fields. 

The sparse set of numerical models so far indicate the following: for fixed $\Rm$, \cite{Schekochihin2007} reported growth rates monotonically decreasing with decreasing $\Pm$ (achieved through increasing $\Rey$) in the range $\Pm \approx 0.1 - 1$. For even smaller values, the growth rates were observed to tend towards a constant value. These findings led \cite{Schekochihin2007} to hypothesize that an asymptotic positive value of the growth rate would exist for high $\Rm$ and low $\Pm$ values. However, such an asymptotic value has not been found yet, even for lower $\Pm$ studies \cite{Warnecke22}.
While, at $\Pm>1$ regime, the growth rates retrieved from numerical experiments are closely consistent with the $\Rm^{1/2}$ scaling expected from Kolmogorov turbulence \cite[reviewed, e.g., by][]{BS05}, the low-$\Pm$ simulations are less consistent with it \cite{ISCM07,Schekochihin2007,Warnecke22}. A better match to a logarithmic scaling law, growth rate being proportional to $\ln \left(\Rm/\Rmc \right)$, valid near the offset of the dynamo action \cite{KR12}, was reported by \cite{Warnecke22}. They concentrated on mapping the region near $\Rmc$, hence this result might not be so unexpected. However, this also does not reveal the true scaling of growth rate, for which simulations far removed from $\Rmc$ would be required. This is an extremely challenging task numerically.

\begin{figure}
\begin{center}
\resizebox{0.95\hsize}{!}{\includegraphics{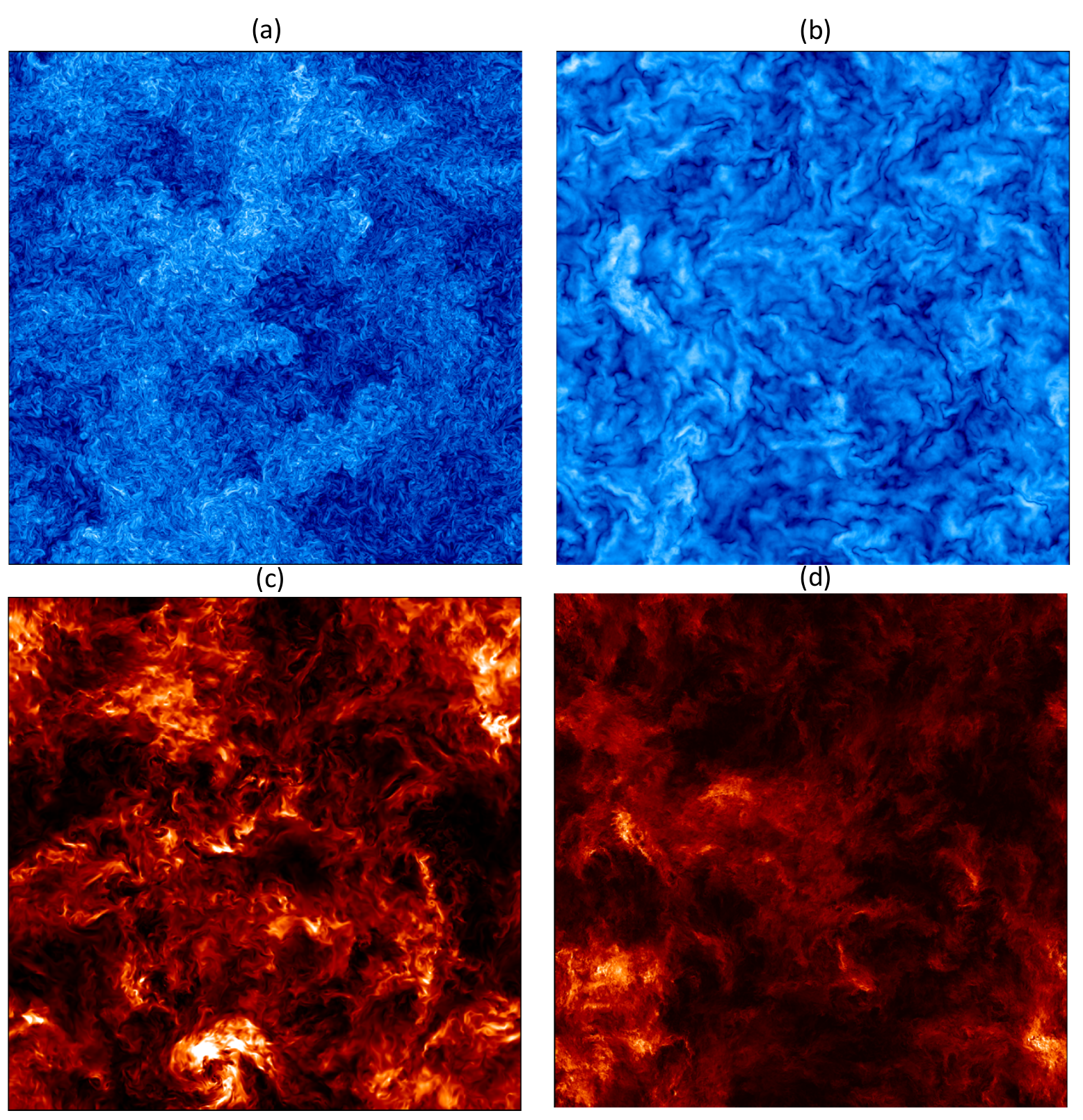}}
\end{center}
\caption{Two-dimensional slices of the magnetic field strength (upper row) and speed (lower row), from $\Pm$=1 simulation (left column) and $\Pm$=0.005 (right column) simulations \cite[reproduced from the models reported in][]{Warnecke22}. The $\Pm$=1 models were run with the resolution of 1024$^3$ and $\Rey=\Rm=4096$, while the low $\Pm$ runs with the resolution of 4096$^3$, $\Rey\approx$ 33,000 and $\Rm$=165. In this model setup, the Reynolds numbers are defined using the forcing wavenumber, and are hence to be multiplied by $2\pi$ to match the definitions used in this paper. The simulation domain is fully periodic and has dimensions of $k_1=2\pi$ in all directions, while the white-in-time plane wave forcing has a mean wave number of $k_f=2k_1$.}\label{fig:lowPm}
\end{figure}

As per the expected magnetic energy spectrum from Kazantsev theory, the simplified low-$\Pm$ models do not yield a direct agreement either \cite{ISCM07,Schekochihin2007,Warnecke22}. Usually, for the numerical convenience, the spectrum is cut short at low wavenumbers, so that the maximum range of higher wavenumbers could be modelled. Hence, in the setups trying to minimize $\Pm$, the expected $k^{3/2}$ scaling of magnetic energy often cannot be seen by design. Nevertheless, at higher wavenumbers between the forcing and resistive scales, the magnetic spectra develop a cascade with negative power laws of varying steepness. \cite{Schekochihin2007} report spectral indices down to -11/3 at low $\Rm$ and of -1 at higher $\Rm$, and \cite{Warnecke22} find a slope approximately of -3 for their simulations near $\Rmc$. As per as the visual appearance 
(see Figure~\ref{fig:lowPm}), the magnetic field exhibits less obvious folded structures having the
width at the resistive scale, 
as would be expected from the Kazantsev model
\cite{ISCM07,Schekochihin2007,Warnecke22}. 
In the case of $\Pm$=1 (left column, panels (a) and (c)), the correlation of magnetic field strength with high/low turbulent speeds is not very strong, while the magnetic field tends not to be volume filling. In the small $\Pm$ case (right column, panels (b) and (d)), the magnetic field has a clearly higher filling factor, and shows even a weaker correlation with the turbulent velocity field. Less folded and thicker structures are seen in the low-$\Pm$ case in comparison to $\Pm$ of unity. 

With turbulent convection, SSD was established for $\Pm > 1$ simulations since the 1990s,
the first successful convection simulation with self-sustaining magnetic fluctuations having been reported by \cite{Nordlund1992}.
Setups with Boussinesq approximation have been confirmed to exhibit SSD action 
\citep[e.g.][]{Cattaneo1999},
and the same applies also to stratified and compressible setups 
\citep[e.g.][]{Nordlund1992,Voegler:Schuessler:2007,Pietarila2010,Hot2015,Bekki2017}. 

Some attempts to go down to $\Pm$s of 0.1 have been performed in local stratified domains \cite{KKB18}
in DNS-like deep convection setups, but no SSD action has been detected at the lower limit. In \cite{KKB18} models, the growth and decay rates at variable $\Pm$ were found to be closely compatible with $\Rm^{1/2}$ scaling, which is different from the behavior seen in the simplified setups described above.
Also, at small wavenumbers, the 3/2 Kazantsev spectrum was not prominent, suggesting that the large-scale motions present in the convection setups (that are energetically too strong in comparison to solar helioseismic observations) also induce excess power to the magnetic energy spectrum, somewhat reminiscent of the phenomenon dubbed the convection 
conundrum \cite[see, e.g., ][]{Hanasoge15}.
In models of solar surface convection with ILES schemes, however, SSD at around $\Pm\approx0.1$ has been found \cite{Rempel:2018,Brandenburg:Rempel:2019}.
In these works a low (numerical) $\Pm$ was realized by combining a more diffusive scheme for the induction equation with a less diffusive scheme for the momentum equation. The resulting effective (numerical) $\Pm$ was estimated from the solution based on the resulting effective diffusivities.

Global or semi-global convection simulations in spherical geometries have been reported to exhibit SSD, albeit still limited to $\Pm=1$ regime \cite{HRY16,KKOWB17,HK21}. In such setups, rotation and stratification are included by default, raising the question whether the fluctuations are genuinely produced by SSD, or rather from a tangling of field generated from a large-scale dynamo \cite[see, e.g., the discussion in ][]{BS05}. The most convincing experiments address this by removing the mean field at each time step, hence allowing for the detection of SSD-generated fluctuations alone \citep[for the method, see e.g.][]{KRB22}. 

In summary, at $\Pm < 1$, the numerical challenges have not yet enabled the detection of SSD in global models of turbulent convection. In the light of the evidence obtained from the more simplified systems, however, the earlier strong doubts about the existence of SSD in turbulent convection at low-$\Pm$ have recently been alleviated. Moreover, this question will be directly addressable in the near future with codes capable of taking advantage of accelerator 
platforms \cite[see, e.g.,][]{Pekkila22,Wright:etal:2021}. 

\subsubsection{Non-linear saturation}\label{subsect:saturation}
Simulating low Pm SSD is an extremely challenging task from a computational power requirement perspective. The simultaneous requirement of high $\Rey$ and an 
$\Rm>\Rmc$ for dynamo action requires extremely high resolutions and long integration times. Hence, even the most
simplified setups operate in the kinematic regime, where the generated magnetic field has negligible back reaction on the flow.
The non-linear regime of the SSD has been mainly studied for $\Pm>1$, while only a handful of studies have been able to address the $\Pm<1$ regime. Non-linear studies, however, are required to draw any conclusions on the effects of the SSD-generated fluctuations on the dynamics of systems like solar and stellar convection zones, and about the interactions of the two dynamo instabilities (namely LSD and SSD).

The study of \cite{Brandenburg:2011:SSD_low_Pm} used DNS-like simulations to investigate SSD in forced non-helical turbulence in the non-linear regime adopting the following strategy. They ran a $\Pm$=1 setup up to saturation, and then kept decreasing the kinematic viscosity while keeping $\Rm$ roughly constant, and continuing the integration from the saturated state with the new parameter values. Their study led to two important findings. Firstly, most of the energy was found out to be dissipated via Joule heating before reaching to the viscous dissipation scale, hence allowing to decrease the kinematic viscosity even further than estimated for the specific grid resolution. Secondly, the saturation strength of the SSD was only weakly dependent on $\Pm$, reducing from roughly 40 percent of equipartition with turbulence to near 10 percent when the $\Pm$ was changed by two orders of magnitude, when computed from volume-averages. However, at all scales, the magnetic energy was sub-dominant to the kinetic energy. Interestingly, the bottleneck effect, dominating the dynamics in the kinematic regime, was suppressed in the nonlinear regime.
An attempt with a similar strategy with turbulent convection was undertaken by \cite{KKB18}, who found that the decrease of the saturation strength of the magnetic field was somewhat, but not detrimentally, stronger with $\Pm$.

\cite{Brandenburg:2014:Pm} and \cite{Sahoo11} performed further idealized simulations in the non-linear regime, reporting on the kinetic and magnetic dissipation rates, and their dependence on $\Pm$. \cite{Brandenburg:2014:Pm} studied both helically and non-helically forced cases, the former also allowing for LSD. 
Interestingly, the ratio of kinetic to magnetic dissipation was observed to exhibit a positive power-law behaviour with $\Pm$ in both scenarios, albeit with somewhat varying power law index in helical versus non-helical cases.
This implies that in the case of low-$\Pm$ dynamos the energy being pumped into the system through the kinetic energy reservoir (be it forcing, convection, shear,...) is converted 
by the Lorentz force
to magnetic energy more efficiently for smaller $\Pm$, and then dissipated through resistive dissipation rather than through the viscous one. 

At first sight these findings appear contradictory, how can the saturation field strength be mostly independent of Pm and the Lorentz force work vary strongly with Pm at the same time? \cite{Brandenburg:Rempel:2019} further studied the SSD
saturation using both forced DNS and convective ILES models and confirmed a similar behavior using both approaches. They further analyzed the transfer function of the Lorentz force in spectral space and found that there is a regime on small scales where the Lorentz force work can be positive, dubbed ``reversed dynamo'', as in the case of a normal dynamo, the flow is doing work against the Lorentz force, and a negative work contribution would be expected. The wavenumber at which this reversed dynamo regime is entered depends strongly on Pm. For a sufficiently low Pm the reversed dynamo is completely absent, but it is growing with increasing Pm until the positive energy transfer of the reversed dynamo on small scales almost completely offsets the negative transfer on large scales. As a consequence a high-Pm dynamo can have in the saturated state a strong field with little net Lorentz force work, while the transfer of energy from kinetic to magnetic energy on large scales remains strong. On small scales this energy is returned to kinetic energy and dissipated through viscosity. 
While the magnetic energy cascade extends to scales smaller than the kinetic energy cascade, there is very little energy left in that cascade. 
In the low-Pm regime the Lorentz force work is negative on all scales and most energy is dissipated through resistivity. 
While the kinetic energy cascade extends to much smaller scales, there is very little energy left in that cascade. As a consequence the saturated dynamo behaves in both cases like a Pm $\sim$ 1 dynamo since kinetic and magnetic energy cascades terminate at a similar scale, which is given by the larger 
of the viscous and resistive dissipation scales.

\subsection{Models of deep convection} \label{subsect:deepconvection}
SSD action has not yet been found in DNS-like models of deep convection with low $\Pm$. This is mostly due to lack of computing resources to properly investigate this regime. There is, 
however, rich literature and exciting findings at $\Pm\geq$1 regime. This regime is either selected on purpose by using explicit viscosity schemes and setting $\nu\geq\eta$, or using ILES schemes, which results in effective $\Pm$s close to unity by inspecting the spectral cut-off scales. 

SSD magnetic fields have been studied in local Cartesian convection setups, to maximize the fluid Reynolds numbers.
In these studies \cite{Hot2015,HRY16,Bekki2017}, an efficient SSD was found to operate, which also resulted in suppression of convective velocities near the base of the convection zone. There, the magnetic and kinetic energies were found to be nearly in equipartition, resulting in the suppression of convective velocities by a factor of two relative to a purely hydrodynamic solution due to the Lorentz force feedback. The enthalpy flux was not, however, observed to be quenched thanks to a simultaneous suppression of horizontal mixing of entropy by the magnetic fluctuations. These results are suggestive of SSD aiding to resolve the convection conundrum by reducing the convective velocities while increasing the convective flux. The work of \cite{Karak18}, motivated by these results, studied cases of
large thermal Prandtl numbers conjectured to be due to the suppression of thermal diffusion by the strong magnetic fluctuations. They could not provide support to these results, however. Therefore, as of writing of this chapter, the issue remains unresolved.

Global and semi-global simulations of solar and stellar magnetism have also recently
reached parameter regimes where SSDs are obtained 
\cite{HRY14,HRY16,KKOWB17,HK21}, but unless rotation is deliberately suppressed as in \cite{HRY14}, LSD cannot be ruled out, and perhaps even then not completely, as anisotropies due to density stratification and inhomogeneties due to boundary conditions would still be present, giving a faint possibility of LSD to be excited.
All these models suggest that
a vigorous SSD would have profound repercussions for the LSD and differential rotation, but the results appear rather divergent and dependent on the viscosity schemes used. For example, \cite{HRY16} report on non-monotonic behavior of the LSD in the presence of SSD - at low resolution and Reynolds numbers with explicit diffusion scheme, the large-scale dynamo and cyclic behavior is obtained, while it gets irregular and sub-critical at medium resolution, when switching to an ILES scheme. Increasing the resolution further, LSD is revived again, attaining saturation strengths larger than in the lowest resolution case. Whether this behavior is due to the change of the explicit/ILES schemes remains unclear, as no such non-monotonicity is observed when explicit schemes are used throughout \cite{KKOWB17}.  

Instead, \cite{KKOWB17} report monotonically increasing values of the mean magnetic field, although the growth clearly slows down, and perhaps tends towards an asymptotic constant value. At the same time the differential rotation is strongly reduced. This can be traced back to the growing small-scale Maxwell stresses which oppose the small-scale Reynolds ones, and hence through that route lead to weaker differential rotation when the SSD becomes more vigorous. This could explain the tendency of the mean magnetic field growth slowing down as function of $\Rm$, hence reflecting only the quenching of the differential rotation rather than any asymptotics. 

In the global magnetoconvection model of \cite{HK21}, on the other hand, increasing the resolution and consequently the Reynolds numbers to higher values than used before in similar type of ILES calculations, again at the effective $\Pm$=1 limit, has been shown to lead to superequipartition of magnetic energy due to SSD at the small scales. This has been observed to result in more solar-like rotation profiles, meaning more radial than cylindrical isocontours of the angular velocity. In these computations, however, no large-scale dynamo has yet been reported to be excited. This could be due to the difficulty of integrating long enough, albeit the simulations extend already up to 4,000 days of solar evolution. 

\subsection{Development of solar surface small-scale dynamo simulations over the past 2 decades}\label{sect:surface}  
Using an incompressible (Bousinesq) convective small-scale dynamo simulation \cite{Cattaneo1999} demonstrated that highly intermittent small-scale field with a saturation field strengths of about $20\%$ equipartition (averaged over the simulation domain) could be reached in a stellar photosphere. This work suggested that substantial fraction of the quiet Sun magnetic field with strength of a few $10$~G could originate from small-scale dynamo action. Later \cite{Bercik:etal:2005} studied turbulent dynamos in solar like (F-M type) stars using anelastic simulations and found comparable results when applied to the Sun. Their work suggested that small-scale dynamos may explain the observed lower limits for X-ray fluxes from solar-like main sequence stars. The first comprehensive small-scale dynamo simulation of the solar photosphere (including compressibility, radiation transport, open bottom boundary conditions and an equation of state accounting for partial ionization) was presented by \cite{Voegler:Schuessler:2007}. The simulation produced a mean vertical magnetic field amplitude of about 30~G in the photosphere and was subsequently compared in detail to observations through forward modeling of spectral lines using both Zeeman \citep{2010A&A...513A...1D} and Hanle \citep{Shchukina:TrujilloBueno:2011} diagnostics. It was found that these simulations fell short by a factor of $2-3$ compared to Hinode observations of Zeeman polarisation in the Fe I $6302\AA$ lines; an even larger discrepancy by an order of magnitude was found comparing the Hanle depolarization in the Sr I $4607\AA$ line, suggesting that in addition to being to weak, the magnetic field was also falling off with height too rapidly. It was found by \cite{Rempel:2014:SSD} that increasing the resolution alone was insufficient to address the discrepancy. A critical component was to account for magnetic field that is transported into the photosphere from the deeper layers of the convection zone (see Sect. \ref{sect:deep_shallow} for further detail). These improved simulations were again compared to observations through forward modeling \citep{Danilovic:etal:2016:spectra,2016A&A...593A..93D,delPinoAleman:etal:2018} and it was found that simulations with a mean vertical field 
strength of around $60-80$~G at optical depth unity in the photosphere were in agreements with constraints from both Zeeman and Hanle diagnostics. \cite{Khomenko:etal:2017} presented small-scale dynamo simulations that included the 
Biermann Battery term \citep{Biermann:1950} in the induction equation. It was found that this term produces at the edge of granules continuously seed fields with a strength of around 
$10^{-6}$~G, which can be amplified by the dynamo to saturation field strength within a few hours of time. While such fields are too weak to make a difference for the saturated dynamo state, this work highlights that fundamental physical processes do provide a lower bound for the quiet Sun magnetic field that is independent from external seeds (e.g. galactic magnetic field amplified during the star formation process).

\subsubsection{Deep versus shallow recirculation} \label{sect:deep_shallow} 
A general challenge of near surface dynamo simulations is the treatment of the bottom boundary. Since closed boundary conditions enforce in the usually adopted shallow domains unphysical recirculation, most photospheric convection setups use bottom boundaries that are open for convective flows and mimic the presence of a deep convection zone. Such an open bottom boundary does make small-scale dynamo simulations ill-posed, since, dependent on the details of the boundary condition, magnetic energy can leave or enter the simulation domain. The work of \cite{Voegler:Schuessler:2007} used boundary conditions that do not allow for a Poynting flux at the bottom boundary. While downflows do transport energy out of the domain (owing to resistive transport right at the boundary), inflows do not transport energy back into the domain. This setup is conservative and demonstrates dynamo action in the presence of little local recirculation and continuous loss of magnetic energy towards the deep convection zone, which was surmised to be a large hurdle for dynamo action in the photosphere \citep{Stein:Review:2003}. As descibed in Section \ref{sect:surface} these models reached a vertical field strength of $30$~G at optical depth unity, which is about a factor of $2-3$ lower than implied by observations.
The deeper parts of the solar convection have a larger magnetic Reynolds [Prandtl] numbers than the photosphere ($10^9$ [$10^{-3}$] instead of $10^5$ [$10^{-5}$]), which should enable small-scale dynamo action over a wide range of scales. Using ILES simulations \cite{Hot2015} found super-equipartition fields near the base of the convection zone in small-scale dynamo simulations of the deep convection zone. In addition, the deeper convection zone will host magnetic field produced by the large-scale dynamo, which may modulate the quiet Sun network field in addition as found in observations \cite{2022A&A...665A.141K}.

A fraction of this field is transported back to the surface and will appear in the photosphere and boost the amplification of field in the surface layers. \cite{Rempel:2014:SSD} captured this effect by allowing for the transport of horizontal field through the bottom boundary and by considering simulations with a closed boundary and complete recirculation and found and increase of the photospheric saturation field strength by about a factor of $2$. Magnetic field that reaches the photosphere from deeper layers (deep recirculation) has undergone substantial horizontal expansion and appears as a rather smooth seed field in the center of granules, while magnetic field being brought back into the photosphere as a consequence of local downflow/upflow mixing (shallow recirculation) appears as a smaller-scale turbulent field at the edge of granules \citep{Rempel:2018}.
 
\subsubsection{Energy transfers, saturation and total power of the dynamo}\label{sect:saturation}

\begin{figure}
\resizebox{\hsize}{!}{\includegraphics{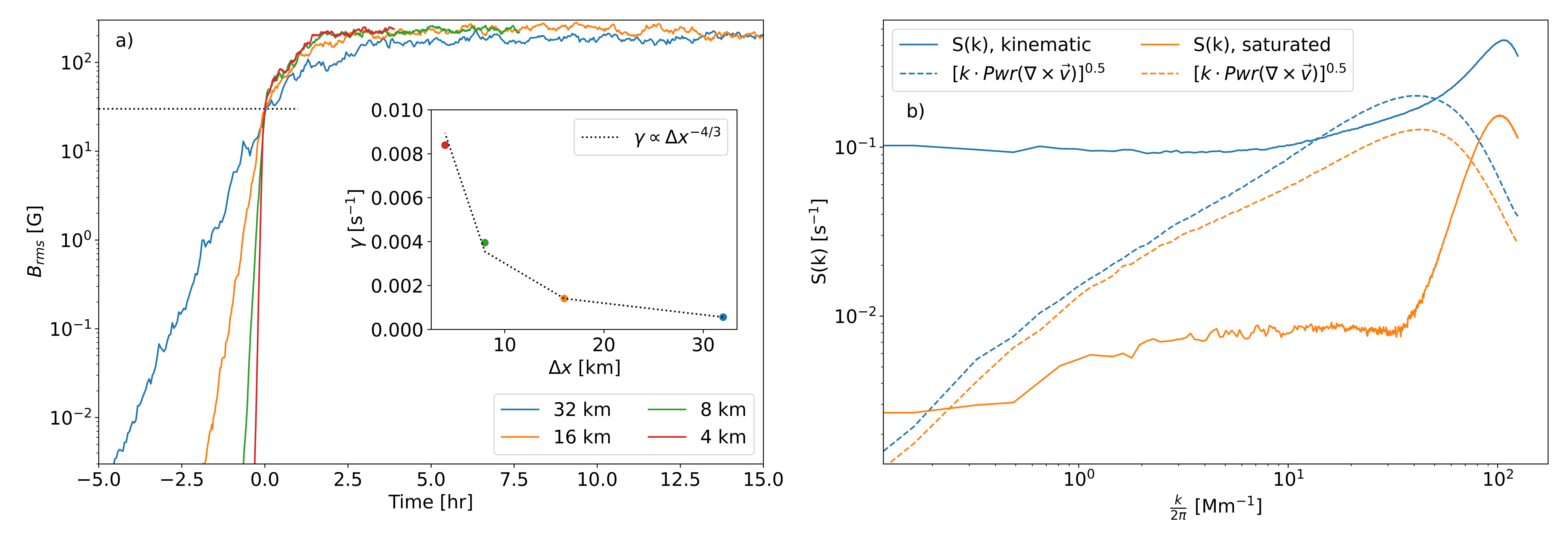}}
\caption{Panel a): Transition from kinematic to saturated phase in 
ILES small-scale dynamo simulations of the solar photosphere for models presented in \cite{Rempel:2014:SSD} with grid spacings from $32$ down to $4$ km. Presented is the time evolution of $B_{\rm rms}$ in the photosphere (optical depth of unity). All models start with the same seed field (around 0.001 G) and the curves are shifted such that the transition from kinematic to saturated phase lines up ($B_{\rm rms}$ of 30 G).
The growth rate during the kinematic phase depends strongly on resolution as shown in the insert. Panel b): Saturation process of a ILES photospheric dynamo simulation. Shown are the effective shear rate (see text) and the vorticity spectrum  during the kinematic and saturated phase.} 
\label{fig:saturation_process}
\end{figure}

Saturation of dynamo action at large magnetic Reynolds numbers is in general not a property of the velocity field, but rather about the relation of velocity and magnetic field to each other. This was demonstrated by \cite{Tilgner:Brandenburg:2008,Cattaneo:Tobias:2009} for both small and large-scale dynamo setups.  They found that the velocity field of the saturated dynamo simulation remains to be an efficient growing dynamo in the kinematic regime, highlighting that saturation is not a property of the velocity field alone and requires a continuous adjustment of the velocity field to small changes in the saturated magnetic field solution. 

Figure \ref{fig:saturation_process}a) shows for various simulations from \cite{Rempel:2014:SSD} the transition from kinematic phase to saturated phase. We show here here $B_{\rm rms}$ in the photosphere, which reaches in the saturated state values around $200-250$ G, corresponding to about $30-40\%$ of the equipartition value.
 While the models show large differences in the their kinematic growth rate depending on their resolution, after passing threshold of about $B_{\rm rms}=30$~G , the remaining slow growth is similar and requires a few hours to reach the final saturation value. The saturation process was further studied in \cite{Rempel:2014:SSD} by looking at the effective shear rate defined in wavenumber space as 
\begin{eqnarray}
	T_{MS}(k)&=&\frac{1}{8\pi}\widehat{\vec{B}}(k)\cdot\widehat{[(\vec{B}\cdot\nabla) \vec{v}]}^*(k)+c.c. \\
	E_M(k)&=&\frac{1}{8\pi}\widehat{\vec{B}}(k)\cdot\widehat{\vec{B}}^*(k)\\
	S(k)&=& T_{MS}(k)/E_M(k)
\end{eqnarray}
During the kinematic growth phase $S(k)$ (Figure \ref{fig:saturation_process}b) has over a wide range of scales a value corresponding to the magnitude of vorticity (given by the quantity $\sqrt{k\cdot \mbox{Pwr}(\nabla\times\vec{v})}$)
on small scales. In the saturated state $S(k)$ has dropped over a wide range of scales by a factor of around $30$ to values comparable to the magnitude of vorticity on the larges scales, while there is only a small reduction of the vorticity by less than a factor of $2$ on the smallest scales. In the saturated state of the dynamo $S(k)$ is small due to combination a misalignment of the shear and magnetic field (reducing $(\vec{B}\cdot\nabla) \vec{v}$) and an induced field being mostly orthogonal to the existing field, minimizing the energy transfer while velocity shear remains mostly unchanged.

The growth rate of the ILES small-scale dynamo shows in the explored range a very strong dependence on resolution (insert in Figure \ref{fig:saturation_process}a) in the form of $\gamma\propto (\Delta x)^{-4/3}$, which is significantly steeper than the scaling of vorticity amplitude $\omega\sim k\cdot v\sim k^{2/3}$ (based on Kolmogorov scaling). For the case with 4km grid spacing the growth rate for magnetic energy (twice the rate of $B_{\rm rms}$) is with $0.017$~s$^{-1}$ about $1/6$ of the rate given by vorticity (Figure \ref{fig:saturation_process}b). It is currently uncertain how this scaling will change at higher resolution.

How much energy is required to maintain the small-scale magnetic field of the Sun? We can provide an estimate based on the previous discussion. With the small value of Pm$\sim$$10^{-5}$ in the photosphere and values not much larger than $10^{-3}$ throughout the convection zone, the dynamo is operating in a regime where the Lorentz force transfers energy from kinetic to magnetic energy on all scales and therefore maximises the power of the dynamo by minimising the energy lost to viscous dissipation. \cite{Rempel:2018} estimated from photospheric ILES dynamo simulations that about $150\,\mbox{erg~cm}^{-3}\, \mbox{s}^{-1}$ are available in the uppermost 1.5 Mm of the convection zone to power the small-scale dynamo, which is integrated over the whole solar surface about $30\%$ of the solar luminosity. The total power of the small-scale dynamo integrated over the volume of the convection zone is bounded by the total pressure buoyancy driving, which is on the order of a few solar luminosities based on mixing length models. We note that the power of the dynamo can exceed a solar luminosity, since it is not a sink of energy. Through Ohmic dissipation this energy is returned to internal energy. If there would be no small-scale dynamo a similar amount of energy would be dissipated through 
viscosity instead. Since the the small-scale dynamo alone is already capable of consuming most of the available convective driving, a large-scale dynamo can only grow at the expense of the small-scale dynamo as it has been suggested by \cite{Cattaneo:Tobias:2014}.
This would imply that the total power of the combined small- and large-scale dynamo does not change much with rotation, while the structuring of magnetic field does change.
Maintaining a magnetic field clumped into sunspots and starspots requires less energy than maintaining a field structured on the smallest scales, which maximises Ohmic dissipation. With a mean vertical magnetic field strength of around 60-80 G in the photosphere, the unsigned flux of the quiet Sun corresponds to about a 100 active regions at any given time. Reorganizing that amount of magnetic flux 
into starspots would turn the sun into a very active star by comparison, while dramatically reducing Ohmic dissipation.  

\subsubsection{Anisotropy of magnetic field in photosphere}\label{sect:anisotropy} 

\begin{figure}
\resizebox{\hsize}{!}{\includegraphics{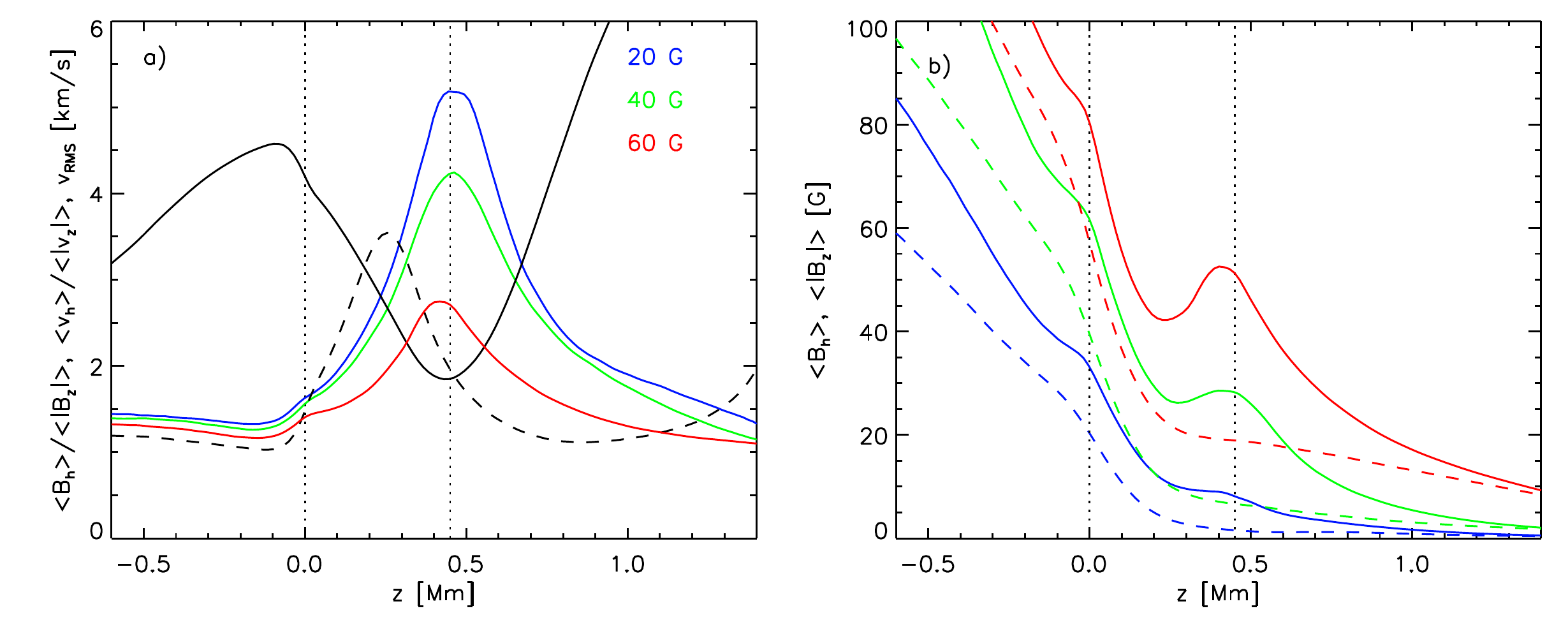}}
\caption{Magnetic field anisotropy in upper convection zone and photosphere. a): Ratio of horizontal to vertical field amplitude for 3 different field strengths (blue, green, red). The black solid line shows the convective RMS velocity, the black dashed line the velocity anisotropy. The peak of magnetic field anisotropy about $450$~km above $\tau_c=1$ coincides with the minimum of convective RMS velocity. b) Height 
variation of vertical (dashed) and horizontal (solid) field amplitudes. The fully saturated dynamo (red line) shows a distinct local peak of horizontal field amplitude at the location of minimum RMS velocity.}
\label{fig:B_anisotropy}
\end{figure}

As summarized in Section \ref{sect:observations}, observations indicate a significant anisotropy of the magnetic field above the photosphere in the sense that the horizontal field components are stronger than the vertical ones. The dynamo simulation of \cite{Voegler:Schuessler:2007} did show a similar preference for horizontal field as reported in \cite{Schuessler:Voegler:2008:bhorz}. It was found that at the height the Hinode lines are sensitive to, the strength of the horizontal field component is 4-5 times stronger than the vertical field component. \cite{Rempel:2014:SSD} analyzed SSD simulation in a wider and deeper domain that had the top boundary located about $1.5$~Mm above the average $\tau_c=1$ level. In these simulations it was found that the ratio of horizontal to vertical field peaks at a height of about $450$~km (see Figure \ref{fig:B_anisotropy}). While the ratio of horizontal to vertical field reached values as high as $5$ during the growth-phase of the dynamo, the ratio dropped to about $2.5$ when the dynamo is saturated (Figure \ref{fig:B_anisotropy}a). Figure \ref{fig:B_anisotropy}b) shows the strength of the vertical and horizontal field components individually. During the kinematic growth phase both drop monotonically with height, however, the vertical field component drops more rapidly with height, which leads to a peak in their ratio at around $450$~km height. During the saturated phase the horizontal field strength does show a distinct peak, while the vertical field strength continues to drop monotonically. However, the magnitude of the magnetic field anisotropy is lower compared to the kinematic phase (a ratio of about $2.5$ instead of $5$). 

What is the origin of the field anisotropy and specifically the origin of the peak in the horizontal field component? Obviously the velocity field is anisotropic above the granulation layer where overturning motions lead to preference of horizontal flows. However, the maximum flow field anisotropy is found at a height of about 250~km  (black dashed line), which is $200$~km lower than the height of peak anisotropy in the magnetic field. The height of peak field anisotropy does coincide with the minimum of the convective RMS velocity (black solid line). This in combination with the distinct peak in horizontal  
field strength may point to the diamagnetic part of turbulent pumping as the mechanism that expels horizontal magnetic field from the photosphere into lower chromosphere where it accumulates in the region with the lowest turbulence intensity.
We note that this explanation is at best qualitative since small-scale dynamo simulations do not have a large scale mean-field, however, the horizontal field overlying the photosphere is organized on scales larger than granules.

Alternative to the approach of using inversions to infer magnetic field anisotropy from observed Stokes profiles, this can also be achieved by analyzing the properties of the Stokes
signals directly, specifically their center-to-limb variation. The simulation highlighted in Figure \ref{fig:B_anisotropy} was compared to Hinode observations by \cite{2017ApJ...835...14L} and a good agreement between the  CLV of synthetic and observed Stokes Q and U was found. This suggests that current photospheric small-scale dynamo  simulations do reproduce to a large extent the observed magnetic field anisotropy. A further test for these models would be multi-height observations that map out the amplitude of horizontal magnetic field in order to test the prediction of a peak in the horizontal magnetic field amplitude at a height of about $450$~km.

\subsubsection{Quiet Sun network field, relevance for coronal heating}\label{sect:network}

\begin{figure}
\resizebox{\hsize}{!}{\includegraphics{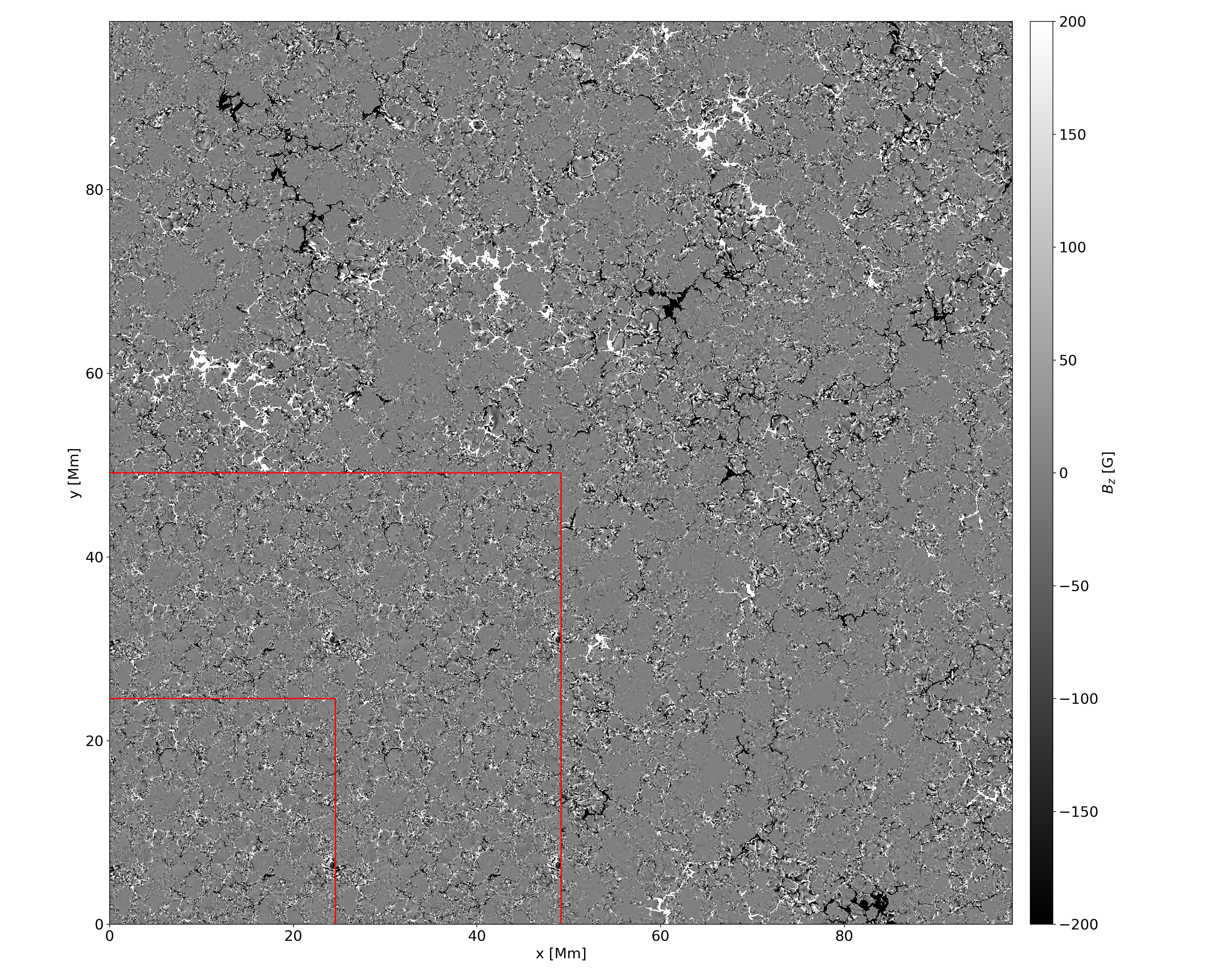}}
\caption{Comparison of $B_z$ at optical depth unity in two dynamo simulations with different domain sizes. The full horizontal extent shows results from a simulation in a $98.304\times 98.304\times 18.432$~Mm$^3$ domain. The small red box indicates a simulation in a $24.576\times 24.576\times 7.68$~Mm$^3$. For better visibility we periodically extended this simulation as a $2\times 2$ tile (large red box).  
The small-scale dynamo simulation in the wider and deeper domain does produce a mixed polarity network structure on scales larger than $10$~Mm, which is absent in the smaller domain.}
\label{fig:ssd_network}
\end{figure}

There are two common misconceptions: firstly, small-scale dynamos can only produce zero-mean fluctuations on the scale of granules and their downflow lanes, and secondly, that large-scale dynamos cannot produce zero-mean small-scale fluctuations through tangling of turbulent motions at the scale of convection. As a consequence both dynamos contribute to the organization of magnetic field on the observable scales from granules to super-granules. Here we focus specifically on the contribution from the small-scale dynamo to network field.

Observations of the quiet Sun during a solar cycle minimum do show a quiet Sun mixed polarity network field \cite[see, e.g., Figure 13 in][]{2008ApJ...672.1237L}), which raises the question of whether this network field is still a remnant of the large-scale dynamo or if it is part of the quiet Sun and maintained by the small-scale dynamo.
Figure~\ref{fig:ssd_network} compares 2 small-scale dynamo simulations, one in a $98.304\times 98.304\times 18.432$~Mm$^3$ domain and one in a $24.576\times 24.576\times 7.68$~Mm$^3$ (see also \cite{Rempel:2014:SSD,Rempel:2020:TSI}). Both simulations were setup with zero netflux on the scale of the simulation domain and both simulations have a mean vertical field strength of about $60$~G. However, in the case of the wider and deeper simulation, we find larger scale flux imbalances that lead to the formation of network field. The larger simulations has on a scale of $24.576\times 24.576$~Mm (extent of small simulation domain) an average flux imbalance corresponding to a mean vertical field of about $10$~G. These flux imbalances are comparable to the imbalance found in the Hinode observations of \cite{2008ApJ...672.1237L} (see \cite{Rempel:2020:TSI} for further discussion).

There are essentially two processes at work that lead to the emergence of a larger-scale network: (1) The dynamo is mostly saturated for the field strengths present in the solar photosphere. While the dynamo is very fast during the kinematic phase, simulations indicate that the kinematic phase ends for field strengths that are about $10\%$ of typical quiet Sun field strengths (see Figure \ref{fig:saturation_process}a). For stronger field the remaining growth time-scale is on the order of several hours, which means that the magnetic field can get organized by photospheric flows on scales larger than granulation. (2) As discussed in Section~\ref{sect:deep_shallow} at least $50\%$ of the small-scale field present in the photosphere originates from deeper layers. 
In a deep, heavily stratified domain the field that is brought up from deeper layers is organized on scales of the deep seated, larger-scale convection and this organization is imprinted on the photosphere and further organized by (1). Further experiments (Rempel, 2022, private comm.) point towards (2) as the critical effect. Setups without deep recirculation (i.e. zero field in upflow regions at the bottom boundary) do not show a network structure even if deeper domains are used.

The quiet Sun network plays a critical role for shaping the upper solar atmosphere. Flux imbalance on larger-scales leads to stronger field reaching the corona, which is in turn critical for maintaining the quiet Sun corona. The need for a small flux imbalance (corresponding to about $5$~G) was identified in models by \cite{Amari:etal:2015:QS_Corona}.
Small-scale dynamo simulations in sufficiently deep domains with deep recirculation naturally produce such a flux imbalance on super-granular scales and have demonstrated that they can maintain a quiet Sun corona at temperatures in the $1-1.5$ million K range \citep{Rempel:2017:MURaM_Corona,Chen:2022:FEM_Corona}.

\subsubsection{Irradiance properties of quiet Sun magnetic field}\label{sect:QA-Irradiance}

While most small-scale magnetic fields in the quiet Sun are too weak to influence the radiative properties of the photosphere, simulations predict a small amount kilo-Gauss strength flux concentrations ($\lesssim 1\%$) in the photosphere which can enhance radiative losses similar to flux concentrations in solar network regions (regions of photosphere with a significant magnetic flux imbalance).  As long as the quiet Sun can be considered as not varying (in the global sense) this will lead to an offset in the total and spectral solar irradiance (TSI/SSI) compared to a hypothetical non-magnetic sun. However, if the quiet Sun varies over solar-cycle or even longer time-scales the quiet Sun magnetic field can make a contribution to the observed variation of TSI and SSI. 
We have to distinguish here between the network and the internetwork
magnetic field. It is known from observations that quiet Sun network does
have some residual variation with the cycle \citep{2022A&A...665A.141K} on the order of 6 G,
and has therefore contributions from both LSD and SSD. It is an open question whether the lowest level of quiet Sun network during the cycle minimum
is a representation of the SSD contribution alone and the therefore the lowest
activity state possible or if a further drop of activity is possible. While observational support for a cycle variation of the internetwork field is marginal (see section \ref{sec:QS_cycle_variation}), we cannot rule it out completely. In order to assess the contribution of quiet Sun field it is necessary to derive the TSI or SSI sensitivity to changes of the quiet Sun field strength. This was investigated by \cite{Rempel:2020:TSI} who computed the TSI and SSI for quiet Sun (zero net flux) and weak network setups and found that a $7$~G change of $\langle\vert B_z\vert\rangle$ on the $\tau_c=1$ level (about $10\%$ variation) causes about a  $0.1\%$ change in TSI. Given that this is about the total observed TSI change over the solar cycle and most of that is explained through contributions from the active Sun, there is very little room for the quiet Sun to vary over the solar cycle (meaning in addition to the quiet Sun network fraction that results from the LSD and is already considered in irradiance models). TSI provides more stringent constraints on quiet Sun variability than direct measurements of the magnetic field \cite[see, e.g.,][] {2014PASJ...66S...4L,2013A&A...555A..33B,Meunier:2018:QS_variation,2021A&A...651A..21F}. While this does not rule out longer-term variations of the quiet Sun, they are very unlikely to happen. The models from \cite{Rempel:2020:TSI}  were used by \cite{Yeo:2020:DimmestSun} to reconstruct solar TSI starting from HMI magnetograms (using radiative MHD simulations to translate HMI magnetograms into irradiance taking into account the full HMI data pipeline). It was found that $97\%$ of the observed TSI variation can be accounted for that way, which gives a significant confidence that radiative MHD simulations capture the radiative properties of magnetic flux concentrations to a significant degree. Using this model they provided a lower bound for a grand minimum irradiance of no more than 2~W/m$^2$ lower than the 2019 solar cycle minimum. To this end SSD simulations were used to represent the lowest activity state of the Sun assuming that only the internetwork field is present. This is a lower bound, since as discussed in Section  \ref{sect:network} a significant fraction of the network field present during a solar minimum could originate from a small-scale dynamo and therefore would be part of the lowest possible activity state as well (i.e. present even during a grand minimum).

\subsection{Radiative zone} \label{subsect:radiative}
The SSD is usually understood to operate in the convection zone, where the field is amplified by turbulent convection. However, no convection is possible in the stably-stratified radiative zones of stars. Does a source of turbulence still exist in these conditions,
and does it have the required strength and properties to drive SSD?
At the tachocline, shear instabilities, gravity waves as well as convective overshoot could, potentially, drive turbulence. Turbulence, in this case, is then no longer isotropic and comes under the realm of stably-stratified systems, an actively researched topic worth its own multiple review articles 
\citep[see, e.g., ][]{ril2012,Lind2006,Cheng2020} and references therein).

The essential feature of such turbulence is the existence of multiple scales over which the characteristics of turbulence change significantly.
The length scales for the largest eddies in the vertical direction (with wavenumbers $k_v$) are much smaller than those in the horizontal direction (with wavenumber $k_h$), i.e., $k_v \gg k_h$. Apart from the usual Reynolds number, another dimensionless number enters into the picture: Froude number ($\rm Fr$). This can be defined in the horizontal and vertical direction in terms of the Brunt-V\"ails\"al\"a frequency $N$ as ${\rm Fr}_h = U k_h/N, {\rm Fr}_v = U k_v/N$. Here, ${\rm Fr_h}$ can be understood as an inverse of the degree of stratification (in stably stratified system, ${\rm Fr}_h<1$) and ${\rm Fr_v}$ can be understood as the ratio of inertial and buoyancy forces. Then there exists the Kolmogorov dissipation wave number $k_{\nu} \sim (\varepsilon/\eta^3)^{1/4}$ \citep{K41} (where the kinetic energy gets dissipated) , the Dougherty-Ozmidov wavenumber $k_O \sim (N^3/\varepsilon)^{1/2}$ \citep{Dou1961,Ozm1965} (where energy in buoyant motions becomes comparable to the kinetic energy), and the buoyancy wavenumber $k_b \sim N/u$ (the length scale corresponding to the adiabatic displacement of a parcel with velocity $u$ in the vertical direction). Under certain assumptions \citep{Lind2006}, these wavenumbers (or, equivalently, length scales) can be described in terms of $\Rey$ and ${\rm Fr}_h$. In addition, in the presence of magnetic fields, there also exists the resistive wavenumber $k_{\eta}$ (where magnetic energy gets dissipated). 

The existence of SSD in solar radiative zone was investigated by \cite{Sko2021} recently. In their paper, they consider a single ${\rm Fr}=u_{\rm rms}/l_i$, where $l_i \sim 1/k_i$ is the integral length scale. Hence, we shall use the same notation to describe their results. They considered 2 situations: i) $k_b<k_{\nu}<k_O$, ii) $k_b<k_O>k_{\nu}$, along with $k_{\eta} \lessgtr k_{\nu}$ for $\Pm \lessgtr 1$. In the first case, the small scale motions down to $k_O$ are suppressed by viscosity making the flow more laminar and unsuitable for SSD growth. For the second case, however, ${\rm Fr}$ comes into the scaling. As $k_{\nu} \sim {\rm Re}^{3/4}\varepsilon^{-1/4}$ and $k_O \sim \varepsilon^{-1/2}{\rm Fr}^{-3/2}$, taking their ratio gives $k_{\nu}/k_O \sim ({\rm Re Fr^2})^{3/4}\varepsilon^{1/4}$. For fixed values of $\Pm$, they explored the $\Rey-{\rm Fr}$ parameter space, and investigated the critical value of $\Rey$ ($\Rey^{c}$) above which SSD action was possible. They found a scaling of $\Rey^c \sim {\rm Fr}^{-2}$ for $\Pm \geq 1$, reducing the parameter space from $\Rey-{\rm Fr}-\Pm$ to ${\rm Rb}-\Pm$, where ${\rm Rb = Re Fr^2}$ is defined as a buoyancy Reynolds number. The two cases can then be distinguished as ${\rm Rb}<1$ and ${\rm Rb}>1$, respectively. Their main result was the existence of a critical ${\rm Rb}$ of around 3 to 9 (for high and low $\Pm$ regime, respectively) above which SSD action was possible. Since SSD fields are expected to be the fastest growing fields, these fields could influence the the operation of instabilities associated with the generation of large scale field at the base of the convection zone \citep{Spr1999}. They also calculated the calculate the critical ${\rm Rb}$ from estimated solar values to be around a 100 (an order of magnitude higher than the critical value), which would imply that such a stably-stratified SSD could, in principle, exist in the solar interior. However, the sensitive dependence of this value on poorly constrained parameters like $u,l_i$ for the solar interior makes this statement somewhat tentative. 

\section{SSD on other cool stars}\label{sect:stellar}
The sun is the only star which we can resolve well enough to study detailed properties of granulation and quiet small-scale solar magnetism. 
For other stars the Zeeman-Doppler imaging inversion technique can be used to study their surface magnetic field \citep[ZDI; see, e.g., ][]{Semel1989}. Due to the lack of surface resolution, this method has the capacity to trace the large-scale magnetic structures only. To estimate the total surface magnetic field, the Zeeman broadening and intensification can be used \citep[see, e.g., ][]{Kochukhov2020}. They demonstrated that only a fraction of the total magnetic field is recovered by ZDI. By combining ZDI with Zeeman broadening and intensification measurements, it is thus possible in principle to estimate how much of the magnetic field is hidden in small-scale structures \citep[][]{2021ApJ...915L..20T}.
Hence  
it is well motivated
to extrapolate our understanding of these phenomena to other stars similar to the sun. In this review we consider the possible impact of this quiet star magnetism from the perspective of observations as well as theoretical modelling. 

\subsection{Structural changes}
As mentioned in earlier sections, most simulations of SSD show the magnetic energy to be within an order of magnitude 
of the kinetic energy. For hotter stars with convection zones (F and earlier spectral types), where surface pressure and gravity are relatively low, the kinetic energy near the surface can also be within an order of magnitude of the internal energy 
(another way to put it would be to say the velocities are near-sonic on average). Because of comparable energies within
these three energy reservoirs, SSD fields could potentially 
affect stratification itself. \cite{Bha2022} showed that 
this is indeed the case for an F3V star, where SSD fields cause a reduction in turbulent pressure to the degree that the near surface stratification changes by $\sim 1\%$ in density and pressure in comparison to a hydrodynamic setup with same parameters. On a more global scale, 
\cite{KKOWB17} and \cite{HK21} showed that SSD fields can 
significantly affect the global-scale dynamics of convection.
In a DNS-like study \cite{KKOWB17} found fluctuating magnetic correlations (so called Maxwell stresses) to grow with the strength of SSD, hence balancing out the velocity correlations (so called Reynolds stresses), both contributing to the angular momentum transport and the generation of differential rotation. In general, in SSD active cases, differential rotation was more difficult to sustain than without it.
In an ILES study \cite{HK21} found superequipartition 
magnetic fluctuations at small scales
deeper in the convection zone, and these
could influence the differential rotation profile itself,
making it better match the solar helioseismic observations. 

\subsection{Basal chromospheric flux}
Here we consider the term "basal" to refer to a minimum level of chromospheric activity that is independent of a stellar magnetic cycle. The contribution of basal chromospheric flux to stellar activity indicators is usually removed to establish better activity-rotation relations and its contribution is empirically estimated from inactive stars \citep{Mit2013}. This basal activity is color-dependent and was initially believed to be due to acoustic heating \citep{Sch1987}. This was the case when it was originally thought that solar basal chromospheric heating is attributable to acoustic waves. However, comparison of observed chromospheric intensity fluctuations to simulations show that high-frequency acoustic waves in non-magnetic regions are not sufficient to account for chromospheric radiative losses \citep{Fos2005}. Recent analyses of flux cancellation in the quiet sun seem to suggest that, at least locally, reconnection of 
IN fields can cause strong temperature increases \citep{Gos2018}. In the absence of sufficiently high-resolution observations, numerous simulations of the solar chromosphere illustrate the importance of magnetic fields in transferring energy to the chromosphere (see section 5 of \cite{Car2019} and the references therein for an overview). In the 
most recent realistic simulations of solar atmosphere with an SSD, Przybylski (2022, in prep.) show that most of the chromospheric heating for the solar case is essentially from torsional motions along field lines of quiet sun fields. With this numerical evidence, it becomes important to account for the effect of SSD fields in analysis of stellar chromospheric activity.

\subsection{Stellar variability and exoplanet detection}
In the last two decades, the field of exoplanet detection has exploded. Dedicated exoplanet hunting missions like Kepler/K2, TESS and, soon, PLATO, in hand with follow-up radial velocity (RV) observations from ground-based echelle spectrometers like CARMENES \citep{carmenes} and ESPRESSO-VLT  (the latter with resolving power over 190,000 in the visible wavelength region) \citep{espresso}, have made it possible to detect and study rocky exoplanets \citep{Pep2013,Rib2018}. The instrumental precision of these modern spectrometers are within 10 cm/s, making it possible, in principle, to detect and characterize an earth-like planet orbiting around a sun-like star (the RV contribution of the earth to the sun is around 9 cm/s).

However, stellar variability remains the single largest source of uncertainty in current observations. Currently, the strategies commonly used to account for stellar variability at small timescales (granulation, p-mode oscillations etc.) revolve around tweaking observation frequency and exposure times \citep{Dum2011} to essentially average them out as best as possible. This can become prohibitively costly to approach noise level of less than 10 cm/s, as granulation is expected to be correlated for timescales much longer than few minutes \citep{Meu2015}. Another approach is to model effects of stellar convection using realistic radiative-MHD simulations of stellar convection to model contributions of granulation to RV signal \citep{Ceg2013}. On the theoretical side, simulations of SSD in solar convection \citep{Hot2015}, as well as near-surface convection for other cool main-sequence stars \citep{Bha2022} have shown a consistent reduction in convective velocities. This reduction in convective velocities may be expected to influence the RV signal of granulation and characteristics of pressure modes.

In addition, the time-averaged stellar photospheric lines show a asymmetry due to granular motions, termed convective blueshift \citep{Dra1987}. This is one of the few observable signatures of stellar granulation and the degree of this blueshift for different spectral lines at different formation heights show a sort of universal scaling with effective temperature for different stellar types \citep{Gra2009,Lie2021}. \cite{Shp2011} showed using a simplified model how convective blueshift could influence the measurement of spin-orbit angles. Bhatia et. al. (2022, in prep) show that in simulations of stellar photospheres with SSD fields, there is a reduction not only in convective velocities but also the scale of granulation (granules appear to be slightly smaller in presence of SSD fields), especially so for hotter spectral types.

Lastly, the center-to-limb variation of spectral intensities (used, among other things, for characterizing exoplanetary transits) is usually estimated from 1D model atmospheres. However, there have been discrepancies in comparison between true limb darkening from exoplanetary transit and model atmosphere based limb darkening \citep{How2011}, which could be accounted for by better models. Existing 3D hydrodynamic simulations already show a systematic difference in limb darkening calculated from 1D atmospheres \citep{Mag2015}. To improve precision of stellar photometry (important for determination of stellar radii as well as for transmission spectroscopy \citep{deWit2013}), it becomes important to account for these effects.

\section{Outlook}
\label{sect:outlook}
Small-scale dynamo simulations that describe the process in solar and stellar convection zones and photosphere require ingredients that go in several aspects beyond the simplified setups that are used to study the fundamentals of the dynamo processes, and their kinematic and non-linear phases, important in their own right.
At a minimum, these setups require stratification and turbulence that is driven through convection, which results in the case of photospheric simulations from a volumetric cooling term based on radiative losses (typically computed from full 3D radiative transfer). Furthermore, the near surface layers of the sun and sun-like stars require an equation of state that accounts for the partial ionization of the most abundant elements. Simulations of the photosphere of the Sun and sun-like stars typically use rather shallow domains that do not reach to the base of the convective envelope and just capture the uppermost $5-10$ density scale-heights. In such setups it is not uncommon to use an open lower boundary condition (allowing for vertical mass and convective energy flux) that mimics the deeper convection zone. Such open boundary conditions lead to a less well determined dynamo problem, since solutions and their saturation will depend on the Poynting flux crossing the lower boundary, specifically the Poynting flux in upflow regions. Such dependence is not unphysical as it describes the not-directly-observable coupling between the photosphere and the deeper layers of the convection zone. Ultimately it will be required to conduct SSD and LSD simulations in domains that reach from photosphere to the base of convection zone to account for the full interaction of all convective scales in the system. Simulations that include all of the above listed ingredients are often referred as ``realistic" or ``comprehensive" simulations, however, their realism is, as in most simulations, determined
by the affordable resolution, which limits the achievable $\Rey$ and $\Rm$ and typically constrains values of $\Pm=\Rm/\Rey$ close to unity. It is not uncommon for these models to use implicit large eddy simulations (ILES) in which diffusion terms are arising from the employed numerical scheme and are typically of a hyper-diffusive nature (higher order than Laplacian) with a non-linear dependence on the solution properties. 

Although in simple cases ILES and DNS-type models 
do produce results that are in good agreement, they tend to disagree in situations where conditions for LSD onset are also fulfilled, that is, where rotation and its non-uniformities together with stratification are also allowed for. This seems to indicate that it does matter for the large-scale dynamics and magnetism how and at which scale the magnetic dissipation takes place, and investigating these issues further is an important future direction in deep convection models. LSD-SSD interactions cannot be ruled out as one decisive mechanism contributing to the convection conundrum and its solution.

The past couple of years have brought important verification steps of SSD excitation both in low-Pm and highly and stably stratified plasmas. While there is now nearly no doubt of the ubiquitous existence of SSD in solar and stellar convection and radiation zones, this raises the need of considering its role in various new scenarios, opening up new, exciting research avenues. 
However, in view of the drastic differences between $\Pm\approx$1 systems and those with low $\Pm$, albeit known mostly only in the kinematic regime, far-reaching conclusions from $\Pm=1$ models should be avoided, and verification at lower $\Pm$ should always be pursued.

Solar observations will continue to play a critical role in
constraining solar dynamo models. 
Of particular importance is to
resolve the question of the isotropy of internetwork fields at
different heights in the atmosphere and to study their temporal
evolution from emergence until disappearance. The spatial distribution
of the emergence sites in the granulation pattern may place additional
constraints on the SSD mechanism and should be investigated
further. DKIST will make it possible to tackle these questions with
unprecedented sensitivity, providing precise linear polarization
measurements over a much larger fraction of the solar surface than has
been possible until now. DKIST will also obtain the first spatially
and temporally resolution Hanle measurements ever, which will open new
avenues for studying the weakest magnetic fields of the quiet solar photosphere. In order to allow for a comparison between simulations and observations we will need in the future higher resolution photospheric SSD simulations that also explore the low Pm regime present in the photosphere. At this point it is unknown if differences between the currently realized Pm$\sim$1 (ILES) simulations and the Pm$\sim$$10^{-5}$ regime of the solar photosphere will be detected or not. 

\backmatter

\bmhead{Acknowledgments}

MJKL acknowledges fruitful discussions with Dr. J\"orn Warnecke and Dr. Thomas Hackman, and funding from the European Research Council (ERC) under the European Union’s Horizon 2020 research and innovation program (Project
UniSDyn, grant agreement no 818665). LBR acknowledges financial support from 
the Spanish MICIN/AEI 10.13039/501100011033 through grants RTI2018-096886-B-C5, PID2021-125325OB-C5, and PCI2022-135009-2, co-funded by "ERDF A way of making Europe", and through the ``Center of Excellence Severo Ochoa'' grant CEX2021-001131-S awarded to Instituto de Astrof\'{\i}sica de Andaluc\'{\i}a. This material is based upon work supported by the National Center for Atmospheric Research, which is a major facility sponsored by the National Science Foundation under Cooperative Agreement No. 1852977. MR received partial funding from NASA grant NNX17AI30G.

\bmhead{Conflict of interests}
The authors declare no conflict of interests.

\bibliography{Maarit,Tanayveer,Matthias,luis}

\end{document}